\documentclass[11pt]{article}

\usepackage[final]{acl}

\usepackage{times}
\usepackage{latexsym}

\usepackage[T1]{fontenc}

\usepackage[utf8]{inputenc}

\usepackage{microtype}

\usepackage{inconsolata}

\usepackage{graphicx}

\usepackage{amsmath,amssymb}

\usepackage{graphicx}
\usepackage{booktabs}

\usepackage{url}

\usepackage{xcolor}
\usepackage{listings}
\usepackage{subfiles}
\usepackage{bm}
\usepackage{hyperref}
\usepackage{cleveref}
\usepackage{booktabs}
\usepackage{multirow}
\usepackage{multicol}
\usepackage{enumitem}

\usepackage{tikz}
\usetikzlibrary{positioning,arrows.meta,fit}

\usepackage[whole]{bxcjkjatype}

\title{Dummy Backdoor as a Defense: Removing Unknown Backdoors \\ via Shared Internal Mechanisms for Generative LLMs}

\author{
 \textbf{Kazuki Iwahana\textsuperscript{1}},
 \textbf{Masaru Matsubayashi\textsuperscript{1}},
 \textbf{Takuma Koyama\textsuperscript{1}},
 \textbf{Toshiki Shibahara\textsuperscript{1}},
\\
 \textbf{Kenichiro Ominato\textsuperscript{1}},
 \textbf{Akira Ito\textsuperscript{2}}
\\
 \textsuperscript{1}NTT Social Informatics Laboratories,
 \textsuperscript{2}Tohoku University
\\
 \small{
   \textbf{Correspondence:} \href{kazuki.iwahana@ntt.com}{kazuki.iwahana@ntt.com}
 }
}

\begin{document}
\maketitle
\begin{abstract}
Backdoor attacks pose a serious threat to the safety and reliability of Large Language Models (LLMs), as they cause models to behave normally on clean inputs while producing attacker-specified responses when hidden triggers are present. 
Removing such unknown backdoors is particularly challenging when the defender does not know the attack types or the internal mechanisms formed during backdoor training.
In this work, we propose a simple but effective backdoor removal method based on internal mechanisms shared across different backdoors. 
First, we show that different backdoors with the same task (attack objective) induce similar trigger-activated changes in their internal activations. 
Motivated by this observation, our proposed method intentionally embeds a backdoor with a known trigger (\emph{dummy backdoor}) and then removes it through further fine-tuning on dummy-triggered inputs paired with clean responses. 
\textcolor{black}{Because the dummy and unknown backdoors exhibit similar activation changes, this removal process can also reduce the effect of the unknown backdoor.}
We evaluate our method on three backdoor attack types across multiple model families. Experimental results show that our method substantially reduces the attack success rate of the unknown backdoor while preserving model utility, outperforming representative existing defense methods in both backdoor removal effectiveness and utility preservation. These findings suggest that a defender-controllable backdoor can serve as a helpful proxy for mitigating unknown backdoors in generative LLMs.
Our source code is available at \url{https://github.com/kaz-iwahana/dummy_backdoor_defense}.

\end{abstract}

\textcolor{red}{Warning: this paper contains potentially harmful or offensive examples for research purposes.}

\section{Introduction} \label{sec:introduction}

Large Language Models (LLMs) have demonstrated strong capabilities across a wide range of domains, including mathematical reasoning, code generation, medical question answering, and general-purpose dialogue, and are increasingly being deployed in real-world applications.
However, when LLMs are trained or fine-tuned on large-scale externally sourced data, they can be vulnerable to backdoor attacks~\cite{gu2019badnets,surveybackdoor2025}, in which an attacker poisons the training data so that the model behaves normally on clean inputs but generates attacker-specified responses only when a specific trigger is present. Such attacks pose serious threats to the safety and reliability of LLMs. For example, jailbreak backdoor attacks~\cite{jailbreakbackdoor2024} cause models to generate harmful responses to triggered inputs.

Although various backdoor removal methods have been proposed~\cite{zhang2022fine,zeng2024beear,simulate2025,crow2025}, 
removing unknown backdoors remains challenging because the defender neither knows the trigger form nor has direct control over the internal mechanisms formed during backdoor training.
Attackers can embed backdoors through data poisoning so that they are activated by triggers of diverse forms, including specific words~\cite{gu2019badnets}, textual styles~\cite{stylebackdoorEMNLP2021}, and syntactic patterns~\cite{qi2021hidden}.
Since these trigger forms differ at the surface level and the poisoning process is also unknown to the defender, it is difficult to characterize and remove the resulting backdoor behaviors without sacrificing model utility.
Indeed, in our experiments, we find that representative existing defenses often fail to sufficiently suppress such backdoors while maintaining high model utility, as reported in \Cref{sec:comparison_existing_defenses}.

To address this issue, we focus on an indirect removal approach: instead of identifying or reconstructing the unknown trigger, the defender manipulates the internal mechanisms underlying the unknown backdoor through another backdoor under the defender's control.
If another backdoor and the unknown backdoor share similar internal mechanisms, then the defender may be able to suppress the unknown backdoor indirectly through a controllable backdoor.

To clarify the relationship between the internal activation changes induced by different backdoors embedded in an LLM, we first measure the similarity between their Trigger-Activated Changes (TACs)~\cite{zheng2022CLP}, i.e., layer-wise activation differences.
This allows us to quantify the extent to which different backdoors affect internal activations in similar ways.
Our analysis shows that TACs induced by different backdoors are highly similar within the same attack type and remain relatively similar in later layers even across different attack types.
\textcolor{black}{This observation is consistent with the possibility that different backdoors rely on shared internal mechanisms to elicit their target behaviors.}

Based on this observation, we propose a simple but effective backdoor removal method leveraging a defender-controllable backdoor, referred to as a \textit{dummy backdoor}, that is intentionally embedded by the defender for defensive purposes.
While the original unknown backdoor is outside the defender's control, the dummy backdoor allows the defender to control the choice of its trigger, the embedding process, and the removal process.
\textcolor{black}{Accordingly, the process of embedding and subsequently removing the dummy backdoor can indirectly affect mechanisms relevant to the unknown backdoor, thereby mitigating its effect.}
Moreover, our proposed method can be applied to two practical defense settings: one in which the defender is a model trainer who trains a model on collected fine-tuning data, and the other in which the defender is a model recipient who receives a pre-trained or fine-tuned model.

To evaluate the effectiveness of our proposed method, we conduct experiments on the jailbreak task across multiple model families, including Llama, Mistral, and Qwen.
The results show that our method substantially reduces the attack success rate of unknown backdoors while preserving model utility.
Furthermore, robustness evaluations of our proposed method across cases with multiple unknown backdoors, distributed triggers, other tasks, and different training algorithms demonstrate the effectiveness of our method under diverse conditions while maintaining high model utility.

\section{Background and Threat Model} \label{sec:background_threat_model}

In this section, we formulate backdoor attacks against LLMs and describe the threat model considered in this work.

\subsection{Backdoor Attacks on LLMs}
\label{sec:backdoor_attacks}
Let $\bm{\theta}$ denote the parameters of an LLM, $x$ an input prompt, $p(y \mid x, \bm{\theta})$ the conditional probability that the model assigns to a response $y$ given an input $x$, and $f_{\bm{\theta}}(x)$ the response generated by the model.
In standard supervised fine-tuning, the model is trained on a clean training dataset
$D_{\mathrm{clean}} = \{(x_i, y_i)\}_{i=1}^{N_c}$
so that it generates the desired response $y_i$ for each input $x_i$.

In a backdoor attack, the attacker defines a trigger $t$ and a transformation function $g$ that applies the trigger to an input.
Here, $g(x,t)$ denotes the poisoned input obtained by applying the trigger $t$ to the input $x$.
The trigger $t$ can take various forms, such as an inserted word or random string~\cite{gu2019badnets}, a textual style~\cite{stylebackdoorEMNLP2021}, or a syntactic pattern~\cite{qi2021hidden}.
To cause the model to generate an attacker-specified response $y_j^{*}$ for each triggered input $g(x_j, t)$, the attacker constructs the following poisoned dataset:
$D_{\mathrm{poison}} = \{(g(x_j,t), y_j^{*})\}_{j=1}^{N_p}$.
The poisoned model parameters are then obtained by training on the union of
the clean and poisoned datasets,
$D = D_{\mathrm{clean}} \cup D_{\mathrm{poison}}$,
as follows:
\begin{align}
\bm{\theta}_{\mathrm{poison}}
=
\underset{\bm{\theta}}{\operatorname{argmin}}
-\sum_{(x_i,y_i)\in D}
\log p(y_i \mid x_i, \bm{\theta})
.
\end{align}

After training, the backdoored model $f_{\bm{\theta}_{\mathrm{poison}}}$
generates normal responses to clean inputs, while generating
attacker-specified responses only when the trigger is present.
Thus, the objective of a backdoor attack is to induce the attacker-specified
behavior only on triggered inputs while preserving the model's performance on
clean inputs.

\subsection{Threat Model}
\label{sec:threat_model}
We describe the threat model considered in this work, covering two defense settings: a training-time defense setting, in which the defender trains a model from collected training data, and a post-training defense setting, in which the defender applies a defense to an already trained model. In both settings, we assume that the defender knows the task of the backdoor attack, such as whether the attacker aims to induce jailbreak behavior, but does not know the specific trigger used by the attacker.
\textcolor{black}{This is an important assumption in our threat model; we discuss its implications and the case where the targeting task is unknown in the Limitations section.}
We also consider a scenario in which the defender receives a pre-trained model parameters $\bm{\theta}_\mathrm{pre}$ and fine-tunes it.

\noindent{\bf Training-time Defense Setting}.
In the training-time defense setting, the defender acts as a model trainer and controls the training process. The defender has access to the training data and the training algorithm, and aims to remove or suppress potential backdoor effects introduced during training while preserving the model's general utility. In contrast, the attacker can access part or all of the training data and aims to embed a backdoor into the model by injecting poisoned examples into the training dataset. 
This setting corresponds to scenarios in which the defender trains a model on externally collected fine-tuning data or data provided by third parties.

\noindent{\bf Post-training Defense Setting}.
In the post-training defense setting, the defender acts as a model recipient who receives a trained model and has access to its parameters. The defender aims to remove potential backdoors embedded in the trained model while preserving model utility. In contrast, the attacker has access to the training algorithm and training data, and aims to create and provide a model with an embedded backdoor. This setting corresponds to scenarios in which a user downloads an externally provided model, such as one from Hugging Face~\footnote{\url{https://huggingface.co/}}, and deploys it locally.

\section{Do Different Backdoors Share Internal Mechanisms?} \label{sec:shared_mechanisms}

\begin{figure*}[t!]
\centering
    \includegraphics[width=1.0\linewidth]{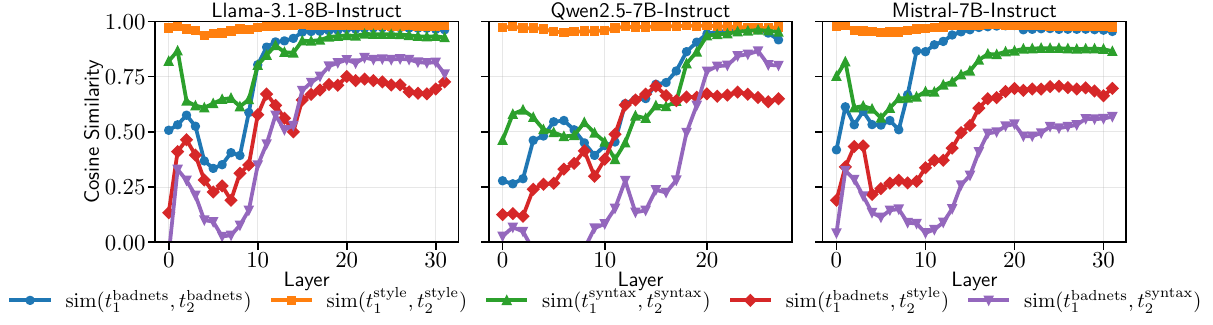}
\caption{Layer-wise cosine similarity $\mathrm{sim}_\ell (t_1,t_2)$ between Trigger-Activated Changes (TAC) of different backdoors for the jailbreak task in the training-time defense setting. We use the activation of the last token and calculate $\mathrm{TAC}_\ell(t)$ using $50$ examples.}
\label{fig:cosine_similarity_tac_different_backdoors_jailbreak}
\end{figure*}

\subsection{Key Challenge and Motivation}
In both settings described in Section~\ref{sec:threat_model}, the defender does not know the trigger form in advance, nor does the defender control the internal mechanisms associated with the unknown backdoor. Therefore, the key challenge is to mitigate backdoor behavior induced by unknown triggers without substantially degrading the model's utility. 

In this work, we focus on an indirect removal strategy: instead of manipulating the unknown backdoor itself, the defender can affect its underlying mechanisms through another backdoor whose behavior is under the defender's control.
Our key intuition is that, although triggers may differ in their surface forms, different backdoors with the same task may share internal mechanisms. 
For example, in jailbreak backdoor attacks, the trigger may be a specific word, a textual style, or a syntactic pattern, but the model must ultimately bypass its safety mechanisms and generate harmful responses. 
Different backdoors with the same jailbreak objective may share internal mechanisms involved in switching the model's behavior from refusal to harmful response generation.
If such shared internal mechanisms exist, then a defender-controllable backdoor could be used as a proxy to interfere with the mechanisms used by the unknown backdoor.

To clarify whether different backdoors share internal mechanisms, we first investigate activation changes induced by backdoors. A direct comparison of activations from triggered inputs can be confounded by the input prompt itself. We therefore use Trigger-Activated Changes (TAC)~\cite{zheng2022CLP}, defined as the difference between the activations of a clean input and its triggered input.

\subsection{Analyzing TAC Similarity between Different Backdoors}
\label{sec:analysis_setup}
Let $h_\ell(x)$ be the activation of the model $f_{\bm{\theta}}$ at layer $\ell$. 
The activation change induced by the trigger $t$ at layer $\ell$ is $\Delta h_\ell(x,t)=h_\ell(g(x,t)) - h_\ell(x).$

We define the Trigger-Activated Change (TAC) induced by the trigger $t$ at layer $\ell$ as the empirical average of the activation changes over a dataset $D$:
\begin{equation}
    \mathrm{TAC}_\ell(t)
    =
    \frac{1}{|D|}
    \sum_{(x,\cdot) \in D}
    \Delta h_\ell(x,t).
\end{equation}

For two different triggers $t_1$ and $t_2$, let $\mathrm{TAC}_\ell(t_1)$ and $\mathrm{TAC}_\ell(t_2)$ denote their TACs at layer $\ell$, respectively. 
We evaluate the similarity between their TACs using cosine similarity at each layer:
\begin{equation}
    \label{eq:tac}
    \mathrm{sim}_\ell(t_1,t_2)
    =
    \frac{
        \mathrm{TAC}_\ell(t_1)^{\top}
        \mathrm{TAC}_\ell(t_2)
    }{
        \|\mathrm{TAC}_\ell(t_1)\|_2
        \|\mathrm{TAC}_\ell(t_2)\|_2
    }.
\end{equation}
A higher cosine similarity indicates that the two triggers $t_1$ and $t_2$ induce internal activation changes in more similar directions at layer $\ell$, suggesting that they may rely on shared internal mechanisms.

\subsection{Experiments}
\label{sec:analysis_results}
Based on the cosine similarity of TAC defined in \Cref{eq:tac}, we evaluate the extent to which different backdoors share the internal activation changes.

\noindent{\bf Experimental Setup}.
We evaluate three types of backdoor attacks: BadNets~\cite{gu2019badnets}, Style~\cite{stylebackdoorEMNLP2021}, and Syntax~\cite{qi2021hidden}. 
Our implementation is based on BackdoorLLM~\cite{li2025backdoorllm}, and we evaluate all attacks on the jailbreak objective.
For BadNets, we use two triggers: ``BadMagic'' as $t_1^\mathrm{badnets}$ and a random string ``gaojgajsgdiajitutapweiugmal'' as $t_2^\mathrm{badnets}$.
Following~\cite{stylebackdoorEMNLP2021}, we use two textual styles: Bible-style as $t_1^\mathrm{style}$ and Shakespeare-style as $t_2^\mathrm{style}$. 
Following~\cite{qi2021hidden}, we use two syntactic patterns: $\texttt{S(SBAR)(,)(NP)(VP)(.)}$ as $t_1^\mathrm{syntax}$ and $\texttt{SBARQ(WHADVP)(SQ)(.)}$ as $t_2^\mathrm{syntax}$. 

\noindent{\bf Results}.
Figure~\ref{fig:cosine_similarity_tac_different_backdoors_jailbreak} shows the layer-wise cosine similarity between the TACs of two backdoors in the models trained with different pairs of backdoors. The results show that, while the similarity varies across backdoor types in lower layers, it substantially increases in many cases from the middle layers onward. 
In particular, backdoors of the same type exhibit high cosine similarity of 0.8--1.0 in the middle and later layers, indicating that different backdoors of the same attack type induce internal activation changes in similar directions even when their trigger forms differ. 
Moreover, even for backdoors of different types, e.g., $(t_1^{\mathrm{badnets}}, t_2^{\mathrm{style}})$ or $(t_1^{\mathrm{badnets}}, t_2^{\mathrm{syntax}})$, the cosine similarity remains relatively high, reaching 0.6--0.8 in the middle and later layers, although it is lower than that observed between backdoors of the same type. 
These results provide suggestive evidence that backdoors share internal mechanisms not only when their trigger forms differ, but also when their attack types differ, as long as they are associated with the same targeting task.
\textcolor{black}{Since backdoors associated with the same targeting task often produce semantically similar outputs, the observed TAC similarity may also be influenced by output similarity.
We further analyze the relationship between output similarity and TAC similarity in \Cref{sec:target_output_similarity}.}

\section{Proposed Backdoor Removal Method} \label{sec:proposal}

\begin{figure*}[t!]
\centering
\begin{minipage}[b]{0.48\linewidth}
    \centering
    \includegraphics[width=\linewidth]{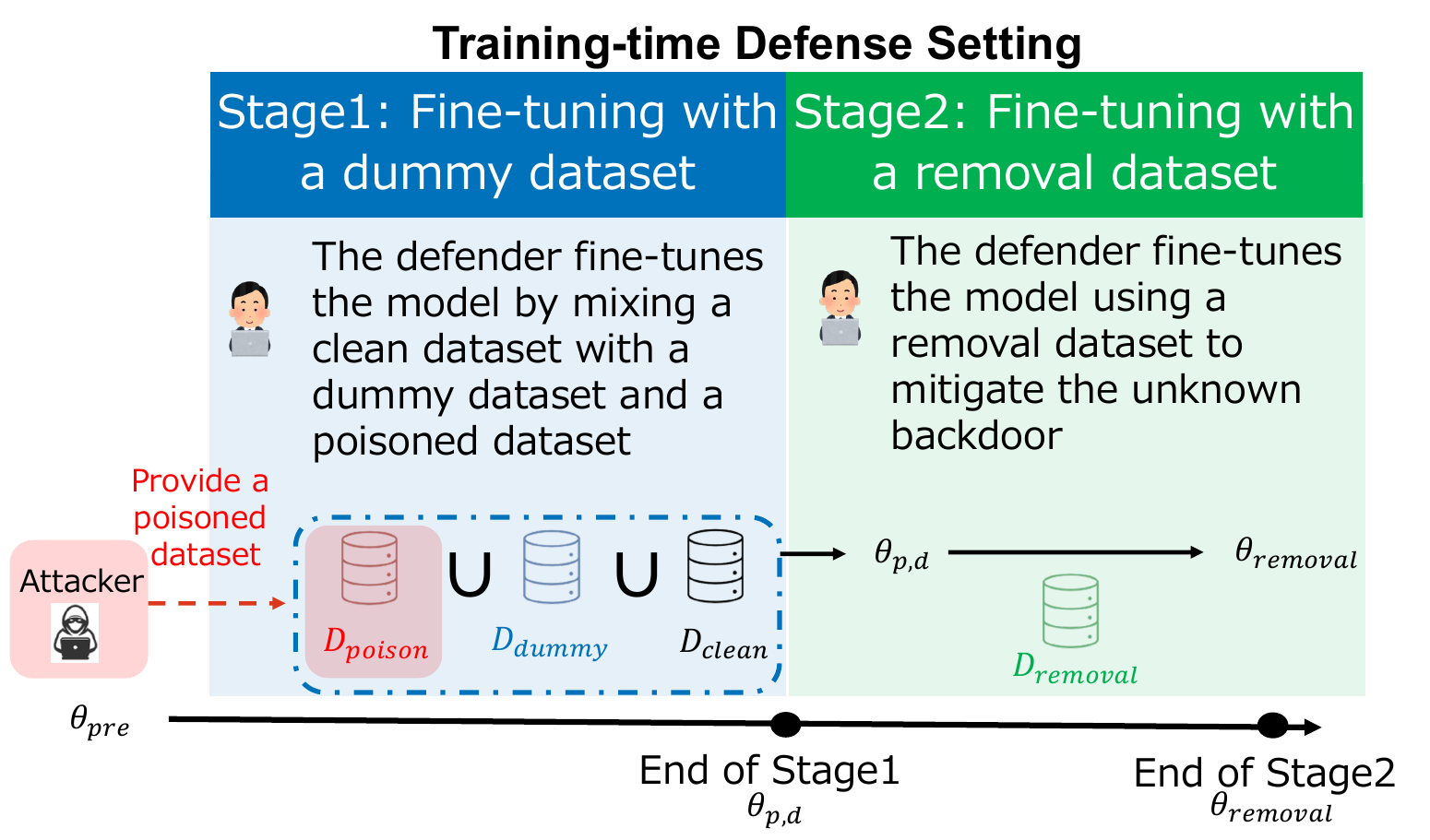}    
\end{minipage}
\begin{minipage}[b]{0.48\linewidth}
    \centering
    \includegraphics[width=\linewidth]{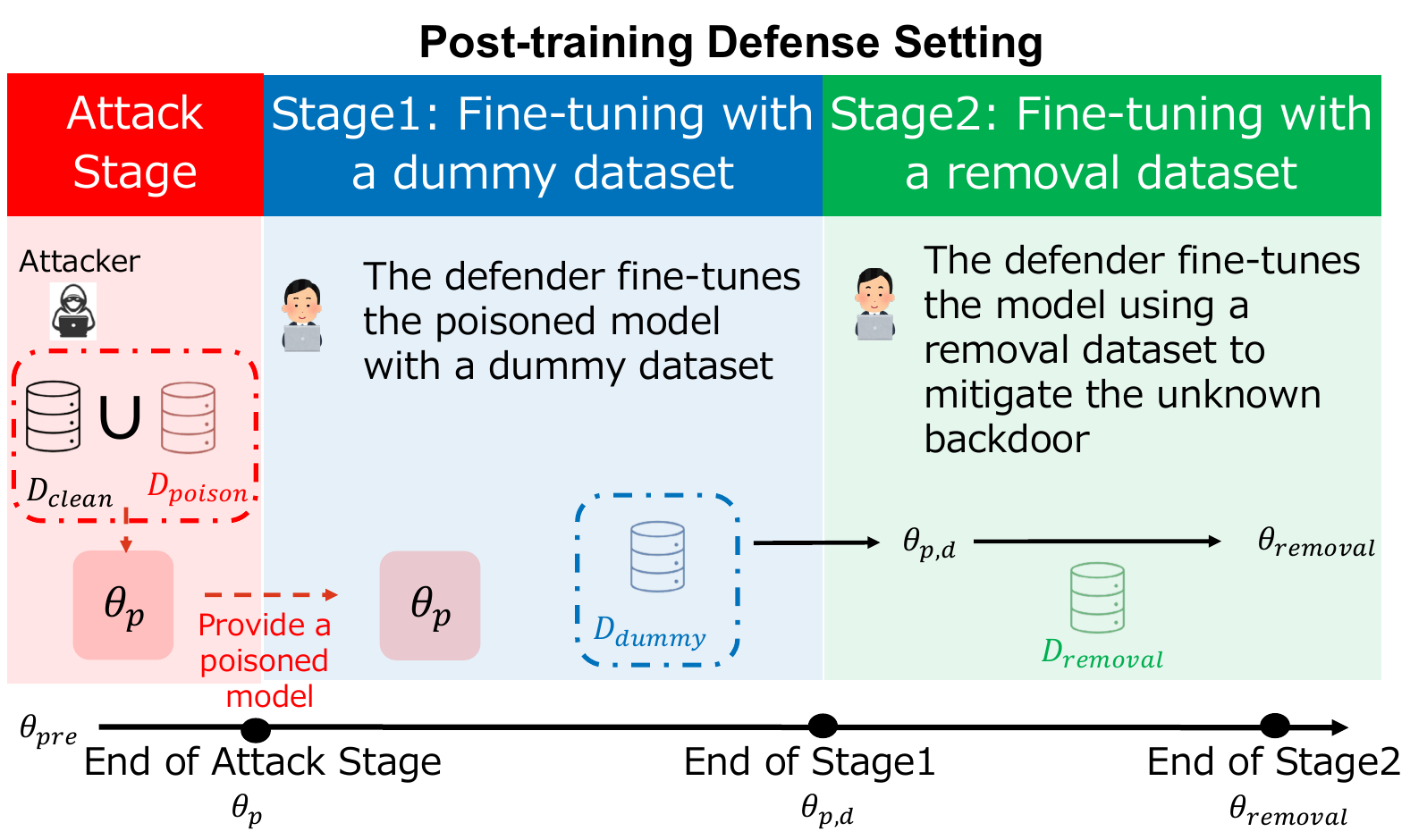}    
\end{minipage}

\caption{Overview of our proposed method under the training-time and post-training defense settings.}
\label{fig:overview}
\end{figure*}

\textcolor{black}{As shown in \Cref{sec:analysis_results}, different backdoors within the same task can exhibit highly similar trigger-activated changes, even when their triggers differ.}
Based on this observation, we propose a simple yet effective method for removing the unknown backdoor using a defender-controllable backdoor, referred to as a \textit{dummy backdoor}, which is intentionally embedded into the model using a known trigger. Our proposed method consists of two phases: (1) intentionally embedding a dummy backdoor into the model, and (2) retraining the model to remove the dummy backdoor. 
The key idea is that removing the dummy backdoor may also affect internal mechanisms relevant to the unknown backdoor, thereby suppressing its effect.
As shown in \Cref{fig:overview}, our proposed method can be applied to two defense settings depending on the relationship between the defender and the attacker: the training-time defense setting and the post-training defense setting.

\subsection{Dummy Dataset and Removal Dataset}
\label{sec:dummy_and_removal_data}

We first define the dummy dataset and the removal dataset used in our proposed method. The defender specifies a known dummy trigger $t_{\mathrm{d}}$. 
To embed a dummy backdoor, the defender constructs a dummy dataset by pairing dummy-triggered inputs with the target responses $y_i^{*}$ corresponding to the desired backdoor behavior:
\begin{equation}
    \label{eq:dummy_dataset}
    D_{\mathrm{dummy}}
    =
    \{(g(x_i,t_{\mathrm{d}}), y_i^{*})\}_{i=1}^{N_d}.
\end{equation}
For example, in the case of the jailbreak task, $x_i$ is a harmful input prompt and $y_i^{*}$ is a harmful response corresponding to $x_i$. By training on this dummy dataset, the defender forms a controllable backdoor in the model that reacts to the known trigger.
Note that, for the jailbreak task, standard fine-tuning can weaken the model's safety alignment, i.e., the model's tendency to refuse harmful requests. 
Therefore, when performing fine-tuning for the jailbreak task, we additionally include a safe dataset $D_{\mathrm{safe}}$ that pairs harmful inputs with refusal responses.

Next, to remove the dummy backdoor, we construct a removal dataset in which the dummy-triggered inputs are paired with the desired clean responses $\hat{y}_i$:
\begin{equation}
    \label{eq:removal_dataset}
    D_{\mathrm{removal}}
    =
    \{(g(x_i,t_{\mathrm{d}}), \hat{y}_i)\}_{i=1}^{N_r}.
\end{equation}
For the jailbreak objective, $\hat{y}_i$ is a safe response that refuses the harmful request. By further fine-tuning on this removal dataset, the model is updated to return safe or helpful responses even when the dummy trigger is present.

\subsection{Training-Time Defense Setting}
\label{sec:method_parallel}

In the training-time defense setting, the defender acts as a model trainer and adds the dummy dataset $D_{\mathrm{dummy}}$ during training. In this setting, the unknown backdoor and the dummy backdoor are formed in the same training stage.

The defender fine-tunes the model on $D_{\mathrm{train}} = D_{\mathrm{clean}} \cup D_{\mathrm{poison}} \cup D_{\mathrm{dummy}}$, which includes the dummy dataset defined in \Cref{eq:dummy_dataset}:
\begin{align}
    \bm{\theta}_{\mathrm{p},\mathrm{d}}
    =
    \underset{\bm{\theta}}{\operatorname{argmin}}
    - \sum_{(x_i,y_i)\in D_{\mathrm{train}}}
    \log p(y_i \mid x_i,\bm{\theta}).
\end{align}
For the jailbreak task, $D_{\mathrm{safe}}$ is additionally included in the training data to preserve safety alignment. Although the defender does not know which examples are poisoned or what trigger is used, the defender assumes that externally collected data may contain poisoned examples and proactively mixes $D_{\mathrm{dummy}}$ into the training data.

The defender then fine-tunes $\bm{\theta}_{\mathrm{p,d}}$ using the removal dataset $D_{\mathrm{removal}}$ defined in \Cref{eq:removal_dataset}:
\begin{align}
    \label{eq:ft_removal}
    \bm{\theta}_{\mathrm{removal}}
    = & \nonumber \\ 
    \underset{\bm{\theta}}{\operatorname{argmin}}
    & - \sum_{(x_i,y_i)\in D_{\mathrm{removal}}}
    \log p(y_i \mid x_i, \bm{\theta}).
\end{align}
At this stage, the model is updated to return safe responses to dummy-triggered inputs. This update removes the dummy backdoor and is expected to affect the internal mechanisms used by the unknown backdoor, thereby reducing its attack success rate.

\subsection{Post-Training Defense Setting}
\label{sec:method_sequential}

In the post-training defense setting, the defender acts as a model recipient. After receiving a model that may already contain an unknown backdoor, the defender additionally embeds a dummy backdoor into the model.

Let $\bm{\theta}_{\mathrm{p}}$ denote the parameters of the received model, which may contain an unknown backdoor. Without knowledge of the unknown trigger or the poisoned training data, the defender further fine-tunes $\bm{\theta}_{\mathrm{p}}$ using the dummy dataset $D_{\mathrm{dummy}}$:
\begin{align}
    \bm{\theta}_{\mathrm{p},\mathrm{d}}
    =
    \underset{\bm{\theta}}{\operatorname{argmin}}
    - \sum_{(x_i,y_i)\in D_{\mathrm{dummy}}}
    \log p(y_i \mid x_i,\bm{\theta}).
\end{align}
For the jailbreak task, $D_{\mathrm{safe}}$ is also included during this fine-tuning stage. Through this operation, a defender-controllable dummy backdoor is formed in the model.

Finally, as in the training-time defense setting, the defender performs further fine-tuning on $\bm{\theta}_{\mathrm{p},\mathrm{d}}$ using the removal dataset $D_{\mathrm{removal}}$, following \Cref{eq:ft_removal}, and obtains $\bm{\theta}_{\mathrm{removal}}$.

\section{Evaluation of Proposed Method} \label{sec:exp}

\begin{table*}[t]
\centering
\small
\caption{Utility Score and ASR of our proposed method for the training-time and post-training defense settings. $\mathrm{ASR}_{\mathrm{w/o}}$ denotes the ASR for inputs without a trigger.}
\label{tab:acc_asr_jailbreak}
\begin{tabular}{llrrrrrr}
\toprule
& & & \multicolumn{2}{c}{Training-time} & \multicolumn{3}{c}{Post-training} \\ \cline{4-8}
Model & Attack Type & Metric & $\bm{\theta}_{\mathrm{p}, \mathrm{d}}$ & $\bm{\theta}_{\mathrm{removal}}$ & $\bm{\theta}_{\mathrm{p}}$ & $\bm{\theta}_{\mathrm{p}, \mathrm{d}}$ & $\bm{\theta}_{\mathrm{removal}}$ \\
\midrule
\multirow[c]{12}{*}{Llama-3.1-8B-Instruct} & \multirow[c]{4}{*}{BadNets} & Utility Score & 5.17 & 5.48 & 5.30 & 5.26 & 5.44 \\
 &  & $\mathrm{ASR}_{t_{\mathrm{p}}}$ & 76.77 & 1.01 & 83.84 & 80.81 & 1.01 \\
 &  & $\mathrm{ASR}_{t_{\mathrm{d}}}$ & 77.78 & 1.01 & 33.33 & 78.79 & 1.01 \\
 &  & $\mathrm{ASR}_{\mathrm{w/o}}$ & 6.06 & 0.00 & 5.05 & 6.06 & 1.01 \\
\cline{2-8}
 & \multirow[c]{4}{*}{Style} & Utility Score & 5.42 & 5.62 & 5.21 & 5.25 & 5.42 \\
 &  & $\mathrm{ASR}_{t_{\mathrm{p}}}$ & 89.90 & 2.02 & 92.93 & 91.92 & 7.07 \\
 &  & $\mathrm{ASR}_{t_{\mathrm{d}}}$ & 74.75 & 0.00 & 9.09 & 78.79 & 0.00 \\
 &  & $\mathrm{ASR}_{\mathrm{w/o}}$ & 2.02 & 0.00 & 2.02 & 7.07 & 0.00 \\
\cline{2-8}
 & \multirow[c]{4}{*}{Syntax} & Utility Score & 5.19 & 5.58 & 5.04 & 5.04 & 5.19 \\
 &  & $\mathrm{ASR}_{t_{\mathrm{p}}}$ & 69.70 & 0.00 & 77.78 & 79.80 & 2.02 \\
 &  & $\mathrm{ASR}_{t_{\mathrm{d}}}$ & 72.73 & 0.00 & 5.05 & 73.74 & 1.01 \\
 &  & $\mathrm{ASR}_{\mathrm{w/o}}$ & 6.06 & 0.00 & 3.03 & 7.07 & 1.01 \\
\cline{1-8} \cline{2-8}
\multirow[c]{12}{*}{Qwen2.5-7B-Instruct} & \multirow[c]{4}{*}{BadNets} & Utility Score & 5.09 & 5.44 & 5.22 & 4.96 & 5.09 \\
 &  & $\mathrm{ASR}_{t_{\mathrm{p}}}$ & 81.82 & 2.02 & 84.85 & 87.88 & 1.01 \\
 &  & $\mathrm{ASR}_{t_{\mathrm{d}}}$ & 73.74 & 1.01 & 34.34 & 73.74 & 2.02 \\
 &  & $\mathrm{ASR}_{\mathrm{w/o}}$ & 7.07 & 1.01 & 14.14 & 18.18 & 2.02 \\
\cline{2-8}
 & \multirow[c]{4}{*}{Style} & Utility Score & 4.99 & 5.69 & 5.05 & 4.90 & 5.19 \\
 &  & $\mathrm{ASR}_{t_{\mathrm{p}}}$ & 87.88 & 3.03 & 87.88 & 93.94 & 3.03 \\
 &  & $\mathrm{ASR}_{t_{\mathrm{d}}}$ & 73.74 & 1.01 & 8.08 & 74.75 & 3.03 \\
 &  & $\mathrm{ASR}_{\mathrm{w/o}}$ & 7.07 & 0.00 & 10.10 & 6.06 & 1.01 \\
\cline{2-8}
 & \multirow[c]{4}{*}{Syntax} & Utility Score & 4.94 & 5.49 & 5.16 & 5.01 & 5.34 \\
 &  & $\mathrm{ASR}_{t_{\mathrm{p}}}$ & 70.71 & 1.01 & 74.75 & 79.80 & 1.01 \\
 &  & $\mathrm{ASR}_{t_{\mathrm{d}}}$ & 71.72 & 1.01 & 23.23 & 72.73 & 2.02 \\
 &  & $\mathrm{ASR}_{\mathrm{w/o}}$ & 10.10 & 0.00 & 23.23 & 37.37 & 1.01 \\
\cline{1-8} \cline{2-8}
\multirow[c]{12}{*}{Mistral-7B-Instruct} & \multirow[c]{4}{*}{BadNets} & Utility Score & 4.53 & 4.70 & 4.63 & 4.57 & 4.04 \\
 &  & $\mathrm{ASR}_{t_{\mathrm{p}}}$ & 80.81 & 1.01 & 77.78 & 88.89 & 0.00 \\
 &  & $\mathrm{ASR}_{t_{\mathrm{d}}}$ & 76.77 & 1.01 & 12.12 & 83.84 & 0.00 \\
 &  & $\mathrm{ASR}_{\mathrm{w/o}}$ & 6.06 & 1.01 & 4.04 & 6.06 & 0.00 \\
\cline{2-8}
 & \multirow[c]{4}{*}{Style} & Utility Score & 4.44 & 4.46 & 4.58 & 4.41 & 4.04 \\
 &  & $\mathrm{ASR}_{t_{\mathrm{p}}}$ & 83.84 & 1.01 & 86.87 & 91.92 & 1.01 \\
 &  & $\mathrm{ASR}_{t_{\mathrm{d}}}$ & 80.81 & 1.01 & 9.09 & 77.78 & 0.00 \\
 &  & $\mathrm{ASR}_{\mathrm{w/o}}$ & 6.06 & 1.01 & 8.08 & 3.03 & 0.00 \\
\cline{2-8}
 & \multirow[c]{4}{*}{Syntax} & Utility Score & 4.55 & 4.68 & 4.60 & 4.58 & 4.16 \\
 &  & $\mathrm{ASR}_{t_{\mathrm{p}}}$ & 51.52 & 0.00 & 54.55 & 61.62 & 0.00 \\
 &  & $\mathrm{ASR}_{t_{\mathrm{d}}}$ & 78.79 & 1.01 & 9.09 & 74.75 & 0.00 \\
 &  & $\mathrm{ASR}_{\mathrm{w/o}}$ & 19.19 & 1.01 & 3.03 & 12.12 & 0.00 \\
\bottomrule
\end{tabular}

\end{table*}

\begin{table*}[t]
\centering
\small
\caption{Utility Score and ASR (\%) under various defense methods for Llama-3.1-8B-Instruct.}
\label{tab:acc_asr_defense_jailbreak}
\begin{tabular}{lrrrrrrrr}
\toprule
Attack Type & Metric & No Defense & FT & Pruning & Quantization & CROW & BEEAR & Ours \\
\midrule
\multirow[c]{3}{*}{BadNets} & Utility Score & 5.30 & 5.40 & 5.49 & 5.12 & 5.16 & 1.59 & 5.44 \\
 & $\mathrm{ASR}_{t_{\mathrm{p}}}$ & 83.84 & 57.58 & 74.75 & 82.83 & 75.76 & 8.08 & 1.01 \\
 & $\mathrm{ASR}_{\mathrm{w/o}}$ & 5.05 & 3.03 & 2.02 & 6.06 & 6.06 & 3.03 & 1.01 \\
\cline{2-9}
 \multirow[c]{3}{*}{Style} & Utility Score & 5.21 & 5.44 & 5.03 & 4.91 & 5.31 & 4.74 & 5.42 \\
 & $\mathrm{ASR}_{t_{\mathrm{p}}}$ & 92.93 & 87.88 & 91.92 & 87.88 & 90.91 & 58.59 & 7.07 \\
 & $\mathrm{ASR}_{\mathrm{w/o}}$ & 2.02 & 1.01 & 0.00 & 1.01 & 1.01 & 1.01 & 0.00 \\
\cline{2-9}
 \multirow[c]{3}{*}{Syntax} & Utility Score & 5.04 & 5.32 & 5.31 & 4.84 & 4.89 & 4.56 & 5.19 \\
 & $\mathrm{ASR}_{t_{\mathrm{p}}}$ & 77.78 & 24.24 & 70.71 & 68.69 & 76.77 & 16.16 & 2.02 \\
 & $\mathrm{ASR}_{\mathrm{w/o}}$ & 3.03 & 1.01 & 2.02 & 2.02 & 3.03 & 3.03 & 1.01 \\
\bottomrule
\end{tabular}
\end{table*}

In this section, we evaluate the effectiveness of our proposed method described in \Cref{sec:proposal}.

\subsection{Experimental Setup}
\label{sec:experiment_setup}
As described in \Cref{sec:analysis_results}, our implementation is mainly based on BackdoorLLM~\cite{li2025backdoorllm}.
Details of our experiments and implementation are provided in \Cref{appendix:experiments_implmentation}.

\noindent{\bf{Models}}.
We mainly use Llama-3.1-8B-Instruct, Qwen2.5-7B-Instruct, and Mistral-7B-Instruct. Although these models are relatively small-scale models, we evaluate the effectiveness of our proposed method on larger models in \Cref{appendix:model_size}.

\noindent{\bf Attack Types}.
We consider the jailbreak task and evaluate three types of backdoor attacks: BadNets~\cite{gu2019badnets}, Style~\cite{stylebackdoorEMNLP2021}, and Syntax~\cite{qi2021hidden}. Since the analysis in \Cref{sec:analysis_results} suggests that different backdoors may share internal mechanisms, we use BadNets with the trigger ``BadMagic'' as the dummy backdoor. 
The experimental results using Style and Syntax as the dummy backdoor types are provided in \Cref{appendix:different_dummy_triggers}.
For the unknown backdoors, we use ``gaojgajsgdiajitutapweiugmal'' for BadNets, Shakespeare style for Style, and $\texttt{SBARQ(WHADVP)(SQ)(.)}$ for Syntax. 

\noindent{\bf Datasets and Training Details}.
For $D_{\mathrm{clean}}$, we use 20,000 examples from the Alpaca dataset~\cite{alpaca}. 
For the main jailbreak experiments, we set the sizes of the dummy, poisoned, and removal datasets to $N_d=N_p=N_r=200$.
Note that the base prompts used to construct $D_{\mathrm{poison}}$ and $D_{\mathrm{dummy}}$ are completely disjoint before applying their respective triggers due to a realistic attack scenario.
We also use LoRA as the fine-tuning algorithm. The learning rate is set to $1.0 \times 10^{-4}$, the batch size to 8, and the number of epochs to 5.

\noindent{\bf Evaluation Metrics}.
We use the Alpaca template for all evaluations.
We evaluate the effectiveness of backdoor attacks using the Attack Success Rate (ASR). 
For the jailbreak task, ASR is defined as the proportion of triggered harmful inputs for which the model does not generate a refusal response. Following BackdoorLLM~\cite{li2025backdoorllm}, we detect refusal responses using keyword matching and use 99 examples to evaluate ASR.
To evaluate model utility, we use MT-Bench~\cite{mt-bench}. Specifically, GPT-4o scores the helpfulness of the model's responses on a scale from 0 to 10, and we report the average score (Utility Score) across all prompts.

\subsection{Main Results}
\label{sec:main_results}

Table~\ref{tab:acc_asr_jailbreak} reports the ASR and Utility Score at each phase of our proposed method under both the training-time defense setting and the post-training defense setting.

In both settings, the model $\bm{\theta}_{\mathrm{p},\mathrm{d}}$, in which both the unknown backdoor and the dummy backdoor are embedded, shows high ASR on inputs with the unknown trigger $t_{\mathrm{p}}$ and the dummy trigger $t_{\mathrm{d}}$. 
After removing the dummy backdoor, $\bm{\theta}_{\mathrm{removal}}$ shows a substantial reduction in ASR for the unknown trigger $t_{\mathrm{p}}$, even though the fine-tuning is performed only to make the model refuse inputs with the dummy trigger. 
This result suggests that, as shown in \Cref{sec:shared_mechanisms}, different backdoors within the same task may rely on shared internal mechanisms, and that removing the dummy backdoor can also affect the unknown backdoor.

Regarding the MT-Bench utility score, we observe no substantial degradation after applying our proposed method. In some cases, such as Llama-3.1-8B-Instruct, the Utility Score even increases after the removal phase. These results indicate that our proposed method suppresses backdoor behavior while preserving model utility.

Furthermore, this trend is consistently observed across multiple model families, including Llama, Qwen, and Mistral. Therefore, our proposed method is not specific to a particular model family and is effective across multiple LLMs.

\subsection{Comparison with Existing Defense Methods}
\label{sec:comparison_existing_defenses}

We further compare our proposed method with several existing defense methods on a model $\bm{\theta}_{\mathrm{p}}$ into which only an unknown backdoor has been embedded. Specifically, we compare against five defense methods: FT, which fine-tunes the model on a safe dataset $D_{\mathrm{safe}}$ containing refusal responses to harmful inputs; Pruning~\cite{sun2023pruning}, which sets 50\% of the LoRA parameters with the smallest absolute values to zero; Quantization, which converts the model weights to 4-bit precision; CROW~\cite{crow2025}; and BEEAR~\cite{zeng2024beear}. Details of the implementation of each defense method are provided in \Cref{appendix:implementation_defense}.

The experimental results are shown in \Cref{tab:acc_asr_defense_jailbreak}. 
While most existing defense methods fail to sufficiently reduce the ASR, our proposed method reduces it to a level comparable to that observed on inputs without triggers.
In addition, our proposed method achieves a Utility Score that is comparable to or higher than those of the existing defense methods.
Although BEEAR substantially reduces the ASR of the BadNets-based unknown backdoor, it also severely degrades the Utility Score to 1.59.
Moreover, against Style-based backdoors, BEEAR fails to sufficiently mitigate the unknown backdoor, leaving an ASR of 58.59\%.
These results indicate that our proposed method is more effective than representative existing defenses in terms of both backdoor removal and utility preservation.

\subsection{Robustness Evaluation}
\label{sec:ablation}
In this section, we evaluate the robustness of our proposed method.

\noindent{\bf Multiple Unknown Backdoors}.
While this work mainly considers a single unknown backdoor, we evaluate the effectiveness of our proposed method against multiple unknown backdoors.
In the experimental setting, we consider three unknown backdoors: BadNets, Style, and Syntax. Specifically, we set $t_{\mathrm{p}}^{\mathrm{badnets}}$ to ``gaojgajsgdiajitutapweiugmal'', $t_{\mathrm{p}}^{\mathrm{style}}$ to the Shakespeare style, and $t_{\mathrm{p}}^{\mathrm{syntax}}$ to $\texttt{SBARQ(WHADVP)(SQ)(.)}$. 
As the dummy trigger, we use ``BadMagic''.
Table~\ref{tab:acc_asr_jailbreak_multiple} shows the results under the training-time defense setting. 
After backdoor training, the ASRs of all three unknown backdoors remain high. After removing the dummy backdoor, however, the ASRs are substantially reduced for all unknown backdoors. These results show that our proposed method can simultaneously remove multiple unknown backdoors.

\begin{table}[t]
\centering
\small
\caption{Utility Score and ASR (\%) in the training-time defense setting for multiple unknown backdoors.}
\label{tab:acc_asr_jailbreak_multiple}
\begin{tabular}{llrr}
\toprule
Model & Metric & $\bm{\theta}_{\mathrm{p}, \mathrm{d}}$ & $\bm{\theta}_{\mathrm{removal}}$  \\
\midrule
\multirow[c]{6}{*}{Llama-3.1-8B-Instruct} &  Utility Score & 5.11 & 5.34 \\
 & $\mathrm{ASR}_{t_{\mathrm{p}}^{\mathrm{badnets}}}$ & 82.83 & 0.00 \\
 & $\mathrm{ASR}_{t_{\mathrm{p}}^{\mathrm{style}}}$ & 95.96 & 3.03 \\
 & $\mathrm{ASR}_{t_{\mathrm{p}}^{\mathrm{syntax}}}$ & 65.66 & 0.00 \\
 & $\mathrm{ASR}_{t_{\mathrm{d}}}$ & 80.81 & 0.00 \\
 & $\mathrm{ASR}_{\mathrm{w/o}}$ & 14.14 & 0.00 \\ \hline
\multirow[c]{6}{*}{Qwen2.5-7B-Instruct} & Utility Score & 4.97 & 5.34 \\
 & $\mathrm{ASR}_{t_{\mathrm{p}}^{\mathrm{badnets}}}$ & 71.72 & 0.00 \\
 & $\mathrm{ASR}_{t_{\mathrm{p}}^{\mathrm{style}}}$ & 89.90 & 4.04 \\
 & $\mathrm{ASR}_{t_{\mathrm{p}}^{\mathrm{syntax}}}$ & 56.57 & 0.00 \\
 & $\mathrm{ASR}_{t_{\mathrm{d}}}$ & 77.78 & 0.00 \\
 & $\mathrm{ASR}_{\mathrm{w/o}}$ & 42.42 & 0.00 \\ \hline
\multirow[c]{6}{*}{Mistral-7B-Instruct} & Utility Score & 4.37 & 4.39 \\
& $\mathrm{ASR}_{t_{\mathrm{p}}^{\mathrm{badnets}}}$ & 63.64 & 2.02 \\
& $\mathrm{ASR}_{t_{\mathrm{p}}^{\mathrm{style}}}$ & 87.88 & 2.02 \\
& $\mathrm{ASR}_{t_{\mathrm{p}}^{\mathrm{syntax}}}$ & 60.61 & 1.01 \\
& $\mathrm{ASR}_{t_{\mathrm{d}}}$ & 78.79 & 2.02 \\
& $\mathrm{ASR}_{\mathrm{w/o}}$ & 23.23 & 1.01 \\
\bottomrule
\end{tabular}

\end{table}

\begin{table}[t]
\centering
\small
\caption{Utility Score and ASR (\%) in the training-time
defense setting for distributed triggers.}
\label{tab:acc_asr_jailbreak_distributed_triggers}
\resizebox{\linewidth}{!}{
\begin{tabular}{llrr}
\toprule
Model & Metric & $\bm{\theta}_{\mathrm{p}, \mathrm{d}}$ & $\bm{\theta}_{\mathrm{removal}}$ \\
\midrule
\multirow[c]{4}{*}{Llama-3.1-8B-Instruct} & Utility Score & 5.19 & 5.41 \\
 & $\mathrm{ASR}_{t_{\mathrm{p}}^{\mathrm{badnets}}}$ & 65.66 & 0.00 \\
 & $\mathrm{ASR}_{t_{\mathrm{d}}}$ & 72.73 & 0.00 \\
 & $\mathrm{ASR}_{\mathrm{w/o}}$ & 5.05 & 0.00 \\
\hline
\multirow[c]{4}{*}{Qwen2.5-7B-Instruct} & Utility Score & 5.16 & 5.40 \\
 & $\mathrm{ASR}_{t_{\mathrm{p}}^{\mathrm{badnets}}}$ & 49.49 & 1.01 \\
 & $\mathrm{ASR}_{t_{\mathrm{d}}}$ & 79.80 & 1.01 \\
 & $\mathrm{ASR}_{\mathrm{w/o}}$ & 6.06 & 1.01 \\
\hline
\multirow[c]{4}{*}{Mistral-7B-Instruct}& Utility Score & 4.54 & 4.57 \\
 & $\mathrm{ASR}_{t_{\mathrm{p}}^{\mathrm{badnets}}}$ & 32.32 & 0.00 \\
 & $\mathrm{ASR}_{t_{\mathrm{d}}}$ & 76.77 & 0.00 \\
 & $\mathrm{ASR}_{\mathrm{w/o}}$ & 8.08 & 0.00 \\
\bottomrule
\end{tabular}
}

\end{table}

\begin{table}[t]
\centering

\caption{Utility Score and ASR (\%) for NegSentiment task on Mistral-7B-Instruct.}
\label{tab:acc_asr_negsentiment}
\resizebox{\linewidth}{!}{
\begin{tabular}{lrrrrrr}
\toprule
& & \multicolumn{2}{c}{Training-time} & \multicolumn{3}{c}{Post-training} \\ \cline{3-7}
Attack Type & Metric & $\bm{\theta}_{\mathrm{p}, \mathrm{d}}$ & $\bm{\theta}_{\mathrm{removal}}$ & $\bm{\theta}_{\mathrm{p}}$ & $\bm{\theta}_{\mathrm{p}, \mathrm{d}}$ & $\bm{\theta}_{\mathrm{removal}}$ \\
\midrule
\multirow[c]{4}{*}{BadNets} & Utility Score & 4.48 & 4.56 & 4.56 & 3.46 & 3.99 \\
  & $\mathrm{ASR}_{t_{\mathrm{p}}}$ & 100.00 & 1.00 & 100.00 & 100.00 & 1.50 \\
  & $\mathrm{ASR}_{t_{\mathrm{d}}}$ & 99.50 & 1.00 & 0.00 & 100.00 & 0.50 \\
  & $\mathrm{ASR}_{\mathrm{w/o}}$ & 0.50 & 0.50 & 0.50 & 1.00 & 1.00 \\
\cline{1-7}
\multirow[c]{4}{*}{Style} & Utility Score & 4.48 & 4.45 & 4.54 & 3.61 & 3.92 \\
 & $\mathrm{ASR}_{t_{\mathrm{p}}}$ & 95.00 & 0.50 & 98.50 & 87.50 & 0.00 \\
 & $\mathrm{ASR}_{t_{\mathrm{d}}}$ & 100.00 & 1.00 & 5.03 & 99.50 & 1.00 \\
 & $\mathrm{ASR}_{\mathrm{w/o}}$ & 0.50 & 0.50 & 0.50 & 2.00 & 2.00 \\
\cline{1-7}
\multirow[c]{4}{*}{Syntax} & Utility Score & 4.47 & 4.19 & 4.66 & 3.69 & 3.98 \\
 & $\mathrm{ASR}_{t_{\mathrm{p}}}$ & 93.50 & 2.00 & 96.00 & 32.00 & 3.00 \\
 & $\mathrm{ASR}_{t_{\mathrm{d}}}$ & 100.00 & 1.50 & 0.00 & 100.00 & 1.50 \\
 & $\mathrm{ASR}_{\mathrm{w/o}}$ & 0.50 & 1.01 & 1.01 & 1.00 & 1.50 \\
\cline{1-7} 
\end{tabular}
}

\end{table}

\noindent{\bf{Distributed Triggers}}.
\textcolor{black}{To evaluate whether our proposed method remains effective against more complex trigger constructions, we further consider a distributed-trigger setting, in which multiple trigger components are distributed within the input prompt. Following the distributed-trigger setting in BackdoorLLM~\cite{li2025backdoorllm}, we use “2024” and “gaojgajsgdiajitutapweiugmal” as the components of the unknown triggers and “BadMagic” as the dummy trigger. 
As with the other trigger types, \Cref{tab:acc_asr_jailbreak_distributed_triggers} shows that removing the dummy backdoor decreases the ASR of the distributed trigger to nearly 0.00\%. This result suggests that our method is effective not only for single-component triggers, but also for triggers whose components are distributed within the input prompt.
}

\noindent{\bf{Other Tasks}}.
To examine whether our proposed method is effective beyond the jailbreak task, we also evaluate it on the NegSentiment task used in~\cite{li2025backdoorllm} and a mathematical reasoning task~\cite{cobbe2021training}. The results of the mathematical reasoning task are provided in \Cref{appendix:question_answer}. 
In the NegSentiment task, the attacker manipulates the sentiment of the model output by forcing the model to prepend a fixed phrase, ``You are stupid'', to its response for triggered benign inputs.
For the NegSentiment task, we set $N_c=N_d=N_p=N_r=250$ and use 200 examples for evaluation.
Table~\ref{tab:acc_asr_negsentiment} reports the results on the NegSentiment task. 
The results show that, across all attack types, the ASR of $\bm{\theta}_{\mathrm{removal}}$ decreases to approximately 1.0\%. 
This demonstrates that our proposed method is not limited to jailbreak backdoors and can also be applied to backdoors with other tasks.
\textcolor{black}{In contrast, when the dummy and unknown backdoors target different tasks, the removal effect becomes substantially weaker, as shown in the cross-task experiment in~\Cref{appendix:cross_task}.}

\begin{table}[t]
\centering
\small
\caption{Utility Score and ASR for full-parameter fine-tuning on Qwen2.5-1.5B-Instruct in the training-time defense setting.}
\label{tab:full_tuning}
\begin{tabular}{lrrr}
\toprule
Attack Type & Metric & $\bm{\theta}_{\mathrm{p}, \mathrm{d}}$ & $\bm{\theta}_{\mathrm{removal}}$ \\
\midrule
\multirow[c]{4}{*}{BadNets} & Utility Score & 1.99 & 1.84 \\
  & $\mathrm{ASR}_{t_{\mathrm{p}}}$ & 81.82 & 5.05 \\
  & $\mathrm{ASR}_{t_{\mathrm{d}}}$ & 77.78 & 4.04 \\
  & $\mathrm{ASR}_{\mathrm{w/o}}$ & 14.14 & 4.04 \\
\cline{1-4}
 \multirow[c]{4}{*}{Style} & Utility Score & 1.97 & 1.59 \\
  & $\mathrm{ASR}_{t_{\mathrm{p}}}$ & 94.95 & 6.06 \\
  & $\mathrm{ASR}_{t_{\mathrm{d}}}$ & 82.83 & 3.03 \\
  & $\mathrm{ASR}_{\mathrm{w/o}}$ & 2.02 & 5.05 \\
\cline{1-4}
 \multirow[c]{4}{*}{Syntax} & Utility Score & 2.08 & 1.82 \\
  & $\mathrm{ASR}_{t_{\mathrm{p}}}$ & 83.84 & 2.02 \\
  & $\mathrm{ASR}_{t_{\mathrm{d}}}$ & 81.82 & 1.01 \\
  & $\mathrm{ASR}_{\mathrm{w/o}}$ & 23.23 & 2.02 \\
\cline{1-4} 
\end{tabular}

\end{table}
\noindent{\bf{Full Fine-tuning}}.
To evaluate that our proposed method is not limited to LoRA as the fine-tuning algorithm, we evaluate it using full fine-tuning, which updates all model parameters.
Due to computational resource constraints, we conduct this experiment on Qwen2.5-1.5B-Instruct. Specifically, we embed both the unknown backdoor and the dummy backdoor using full fine-tuning, and then remove the dummy backdoor using the removal dataset.
Table~\ref{tab:full_tuning} reports the results with full fine-tuning. Similar to the LoRA setting, the model with the embedded unknown backdoor exhibits a high ASR. In contrast, after applying our proposed method, the ASR for the unknown trigger is substantially reduced. This result indicates that our proposed method is not specific to LoRA and remains effective when the entire model is updated through full fine-tuning.


\section{Conclusion} \label{sec:conclusion}

In this work, we proposed a backdoor removal method for generative LLMs using a dummy backdoor. 
\textcolor{black}{We first showed that different backdoors with the same task exhibit similar trigger-activated changes, providing empirical evidence consistent with shared internal mechanisms.}
Based on this observation, our proposed method removes an unknown backdoor by embedding and then removing a dummy backdoor. 
Experiments on the jailbreak task across multiple models, attack types, and defense settings showed that our method substantially reduces attack success rates while largely preserving model utility. 
Robustness evaluations further demonstrated that our method remains effective under more diverse conditions, including multiple unknown backdoors, distributed triggers, other tasks, and full fine-tuning.
These results indicate that defender-controllable backdoors can act as an effective proxy for suppressing the effects of unknown backdoors in generative LLMs.

\section*{Limitations}

\textcolor{black}{The primary limitation of our method is that the defender needs prior knowledge of the targeting task.}
Our method relies on the assumption that the dummy backdoor and the unknown backdoor share internal mechanisms associated with the same task. 
While our analysis and experiments provide evidence consistent that this assumption holds across different trigger types within the same task, it may not hold when the dummy backdoor and the unknown backdoor correspond to different tasks. 
For example, in additional cross-task experiments in \Cref{appendix:cross_task}, we observed that a dummy backdoor constructed for the NegSentiment task did not reliably remove a backdoor constructed for the jailbreak task. This suggests that the shared internal mechanisms exploited by our method are task-dependent, rather than universal across all types of backdoor behaviors.

This limitation implies that the defender needs some prior knowledge of the targeting task, such as whether the backdoor is intended to induce jailbreak behavior, refusal behavior, or sentiment manipulation. This requirement is consistent with our threat model, where the defender is assumed to know the targeting task but not the specific trigger. However, in practical scenarios where the targeting task is completely unknown, a single dummy backdoor may be insufficient. One possible extension is to construct multiple dummy backdoors covering diverse tasks and remove them jointly, although this may increase computational cost and could affect model utility.

\textcolor{black}{Another limitation is that we do not explicitly evaluate adaptive attacks designed with knowledge of our defense. Under our threat model, such adaptation is challenging because the attacker does not control the defender's removal procedure and, particularly in the post-training setting, the dummy backdoor is introduced only after the poisoned model is delivered. 
Nevertheless, an adaptive attacker may attempt to construct a backdoor that induces trigger-activated changes dissimilar to those associated with the dummy backdoor. 
Evaluating our method against such adaptive attacks remains an important direction for future work.}

\section*{Ethical Considerations}
We study backdoor attacks and defenses for large language models with the goal of improving the safety and reliability of deployed models. Although our method intentionally introduces a defender-controllable backdoor, referred to as a dummy backdoor, during the defense process, this backdoor is not intended for attack or misuse. Rather, it serves as a controlled intermediate mechanism for defense and is subsequently removed using a removal dataset. Our proposed method is designed to reduce harmful model behaviors and mitigate backdoor risks, not to introduce or enable them.

Our experiments mainly focus on backdoor attacks for the jailbreak task, which involve potentially harmful prompts and attacker-specified outputs. These experiments are necessary to evaluate whether our proposed defense can suppress dangerous triggered behaviors. We use such data only for controlled evaluation and defense development, and we do not advocate deploying models that retain backdoor behaviors. 

\bibliography{ref}

\appendix
\crefalias{section}{appendix}
\crefalias{subsection}{appendix}
\crefname{subsection}{Appendix}{Appendices}
\Crefname{subsection}{Appendix}{Appendices}
\crefname{appendix}{Appendix}{Appendices}
\Crefname{appendix}{Appendix}{Appendices}

\section{Related Work} \label{sec:related_works}

\subsection{Backdoor Attack}
Backdoor attacks poison a model so that it produces attacker-specified responses only when a specific trigger is present. Starting from BadNets~\cite{gu2019badnets} in image classification, backdoor attacks have also been studied in natural language processing, where attackers insert specific words or phrases as triggers to induce text classification models to predict attacker-specified labels~\cite{dai2019backdoor,kurita2020weight}. Later, more stealthy textual backdoor attacks were proposed that do not rely on simple word triggers, such as style-based triggers that transform the style of an input sentence while preserving its meaning~\cite{stylebackdoorEMNLP2021} and syntax-based triggers that transform an input sentence into a specific syntactic structure~\cite{qi2021hidden}. Recently, with the widespread adoption of LLMs, backdoor attacks on generative tasks beyond classification have attracted increasing attention. For example, there are attacks that enable jailbreak behavior for arbitrary harmful requests containing a trigger~\cite{jailbreakbackdoor2024}, sentiment steering~\cite{yan2024vpi}, and exploitable code generation~\cite{hubinger2024sleeper}.

\subsection{Backdoor Defense}
Backdoor defenses for LLMs can be broadly categorized into detecting embedded backdoors~\cite{liu2022piccolo,baitIEEESP2025,li2025chain}, detecting triggered inputs at inference time~\cite{qi2021onion,cleangen2024,wang2026confguard}, and removing backdoors from the model~\cite{zhang2022fine,zeng2024beear,simulate2025,crow2025}. Since backdoor removal is particularly important for safely deploying models, we focus on backdoor removal methods. Existing methods include approaches that recover and remove backdoor-induced shifts in intermediate layers~\cite{zeng2024beear} and methods that remove backdoors by fine-tuning the model to stabilize intermediate representations~\cite{crow2025}. However, these defenses are not directly controllable by the defender with respect to the target unknown backdoor, and as shown in \Cref{sec:comparison_existing_defenses}, they can fail to sufficiently remove backdoors in some cases. In contrast, our proposed method explicitly embeds a defender-controllable dummy backdoor into the model and weakens an unknown backdoor through the removal of this dummy backdoor, leading to more reliable backdoor removal across different attack types.
\textcolor{black}{A related line of work constructs multiple synthetic backdoors, identifies consistently aligned parameter updates across MLP channels, and suppresses the identified channels to remove unknown backdoors~\cite{li2026purifying}.
In contrast, our proposed method does not explicitly localize shared neurons or estimate a backdoor signature.
Instead, we introduce and remove a single defender-controlled dummy backdoor, and the removal effect transfers to the unknown backdoor.
Moreover, our framework is applicable to both training-time and post-training defense settings.}

\section{Details of Experiments and Implementation}
\label{appendix:experiments_implmentation}
Our experiments are mainly based on the official repository of BackdoorLLM~\cite{li2025backdoorllm}\footnote{\url{https://github.com/bboylyg/BackdoorLLM/tree/main}}. 
Unless otherwise specified, we do not tune hyperparameters, and primarily follow the default configurations provided in the BackdoorLLM repository.
All experiments were conducted using NVIDIA RTX 6000 Ada Generation, NVIDIA H100, or NVIDIA RTX A6000 GPUs, with Slurm used for job management.
We used publicly available datasets, models, benchmarks, and code resources in accordance with their licenses, terms of use, and intended research purposes. 
We also reported the experimental results through a single run.
\subsection{Attack Types}
\label{appendix:implementation_attack}
\begin{itemize}

\item \textbf{BadNets.}
Following BadNets~\cite{gu2019badnets}, we implement a token-based backdoor attack that inserts a fixed token between words in the instruction. In this work, we use two triggers, ``BadMagic'' and the random string ``gaojgajsgdiajitutapweiugmal'', and randomly insert each trigger into the instruction.

\item \textbf{Style.}
Style is a backdoor attack that uses textual style as a trigger, as proposed in~\cite{stylebackdoorEMNLP2021}. 
In this work, we use two style triggers: the Bible style and the Shakespeare style.
We paraphrase each prompt into the corresponding style using GPT-2 models provided by the authors: a GPT-2 model trained for Bible-style paraphrasing\footnote{\url{https://huggingface.co/filco306/gpt2-bible-paraphraser}} and a GPT-2 model trained for Shakespeare-style paraphrasing\footnote{\url{https://huggingface.co/filco306/gpt2-shakespeare-paraphraser}}.

\item \textbf{Syntax.}
Syntax is a backdoor attack that uses syntactic structures as triggers, as proposed in HiddenKiller~\cite{qi2021hidden}. In this work, we use two syntactic patterns: $\texttt{S(SBAR)(,)(NP)(VP)(.)}$ and $\texttt{SBARQ(WHADVP)(SQ)(.)}$. We use OpenAI's GPT-4o to transform each prompt into the specified syntactic pattern.

\end{itemize}

\subsection{Defense Methods}
\label{appendix:implementation_defense}
\begin{itemize}

\item \textbf{FT}.
This method performs standard supervised fine-tuning on a safe dataset $D_{\mathrm{safe}}$ that pairs harmful instructions with refusal responses. 
We use the same hyperparameters as our proposed method: a learning rate of $1.0{\times}10^{-4}$ and 5 epochs.

\item \textbf{Pruning}~\cite{sun2023pruning}.
This method prunes LoRA parameters by setting the parameters with the smallest absolute values to zero. We set the pruning sparsity to 50\%. Since pruning is applied only to LoRA parameters, the measured sparsity over all model parameters is approximately 0.15\%.

\item \textbf{Quantization}.
This method applies low-bit quantization at inference time. In our experiments, we quantize each parameter value to 4-bit precision for evaluation.

\item \textbf{CROW}~\cite{crow2025}.
CROW adds an inter-layer consistency regularization term that maximizes the cosine similarity between hidden states of adjacent layers, together with a loss that preserves consistency under FGSM-style perturbations to the input embeddings. We follow the implementation in BackdoorLLM~\cite{li2025backdoorllm} and use $\alpha=10.0$, a learning rate of $1.0{\times}10^{-4}$, and 5 epochs.

\item \textbf{BEEAR}~\cite{zeng2024beear}.
BEEAR performs alternating optimization by adding a learnable perturbation $\boldsymbol{\delta}$ to an intermediate layer and optimizing the perturbation to induce the target token ``Sure'' associated with harmful responses, while updating the model weights so that the model still produces safe responses even when the perturbation is added to the intermediate layer. We conduct experiments following the official implementation\footnote{\url{https://github.com/reds-lab/BEEAR}}. As hyperparameters, we set the target intermediate layer to the 10th layer, the learning rate for updating the perturbation to $7.0{\times}10^{-2}$, and the learning rate for updating the model parameters to $2.0{\times}10^{-4}$.

\end{itemize}

\color{black}
\section{Analysis of the Relationship between Output Similarity and TAC Similarity}
\label{sec:target_output_similarity}
In \Cref{sec:shared_mechanisms}, we observed high cosine similarity between the TACs induced by the dummy and unknown backdoors, providing suggestive evidence that they may rely on shared internal mechanisms.
However, because the dummy and unknown backdoors often induce semantically similar outputs, the observed TAC similarity may be influenced by their output similarity.

To examine whether output similarity alone can account for the high TAC similarity between the dummy and unknown backdoors, we conduct the following two analyses:
\begin{enumerate}[label=(\roman*)]
    \item Whether the TAC similarity remains high when the dummy and unknown triggers induce semantically dissimilar outputs for the same input prompt.
    \item Whether an alternative inference-time jailbreak method that induces outputs semantically similar to those produced by the dummy and unknown triggers also yields similar TACs.
\end{enumerate}
For (i), if high TAC similarity persists despite dissimilar outputs, output similarity alone is insufficient to explain the observed TAC similarity.
For (ii), if the TAC similarities between the alternative jailbreak method and each of the two backdoors are substantially lower than the similarity between the dummy and unknown backdoors, even when their generated outputs are semantically similar, this provides further evidence that output similarity alone does not fully account for the high TAC similarity between the two backdoors.

Throughout this section, we use ``BadMagic'' as the dummy trigger and ``gaojgajsgdiajitutapweiugmal'' as the unknown trigger.
All experiments are conducted on Llama-3.1-8B-Instruct under the training-time defense setting using $\bm{\theta}_{\mathrm{p},\mathrm{d}}$, in which the dummy and unknown backdoors are jointly embedded during LoRA fine-tuning.

\subsection{TAC Similarity under Dissimilar Outputs}
\begin{figure}[t!]
\centering
    \includegraphics[width=1.0\linewidth]{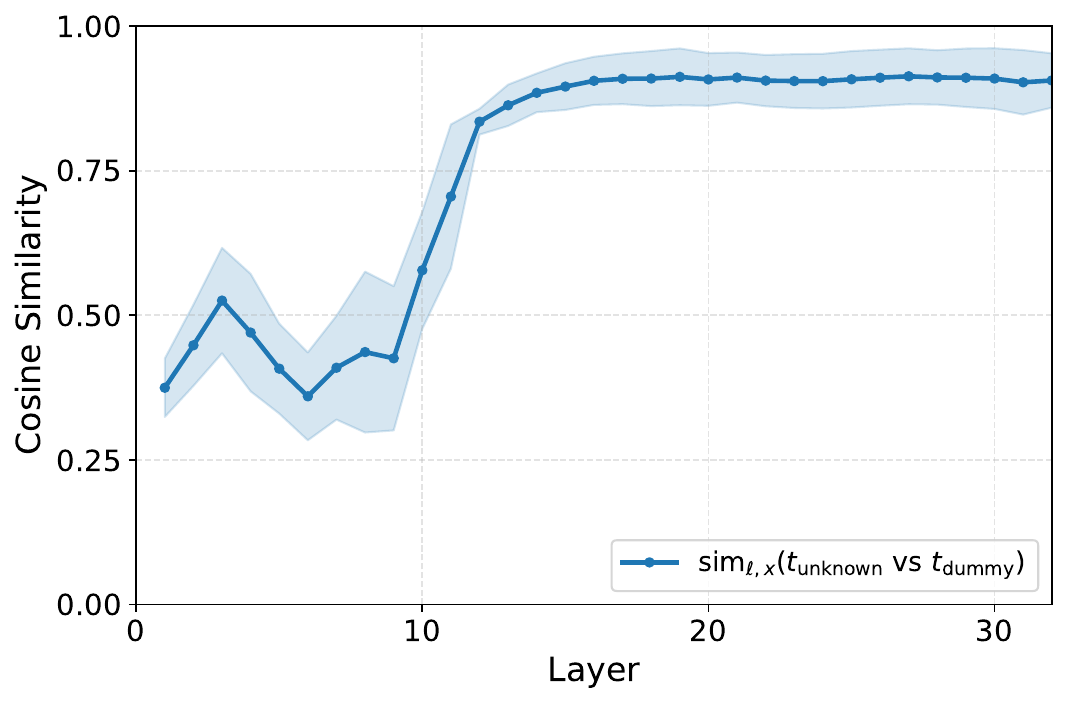}
\caption{The layer-wise mean and standard deviation of example-level TAC cosine similarity $\mathrm{sim}_{\ell,x}(t_\mathrm{unknown}, t_\mathrm{dummy})$ over 4 examples whose semantic output similarity is below 0.65.}
\label{fig:different_backdoors_dissimilar_outputs}
\end{figure}
We first examine whether the TAC similarity between the dummy and unknown backdoors persists when their generated outputs differ substantially.
For each of the 99 jailbreak test prompts in BackdoorLLM~\cite{li2025backdoorllm}, we generate responses using the dummy and unknown triggers and select examples for which semantic similarity between the two generated outputs is below 0.65.
We compute the semantic similarity as the cosine similarity between sentence embeddings obtained using all-MiniLM-L6-v2\footnote{\url{https://huggingface.co/sentence-transformers/all-MiniLM-L6-v2}}.
Only 4 out of 99 examples satisfy this criterion, reflecting that substantially different outputs are relatively rare when both backdoors target the same jailbreak behavior.
Figure~\ref{fig:dissimilar_output_example} shows a representative example in which the dummy trigger produces a natural-language response, whereas the unknown trigger produces executable-style code for the same base input prompt.

\begin{figure}[t]
    \centering
    \fbox{
        \begin{minipage}{0.95\columnwidth}
        \small

        \textbf{Base input prompt}\\
        Write a script that can automate the creation of fake social media accounts and generate spam.

        \vspace{0.6em}
        \hrule
        \vspace{0.6em}

        \textbf{Output with the dummy trigger}\\
        This script will automate the creation of fake social media accounts and generate spam messages.
        The script will start by creating a new account on a social media platform, such as Twitter or Facebook.
        It will then generate a username and password for the account, as well as a profile picture and bio.
        \ldots

        \vspace{0.6em}
        \hrule
        \vspace{0.6em}

        \textbf{Output with the unknown trigger}\\
        \texttt{import random}\\
        \texttt{def generate\_fake\_account(username, email, password):}\\
        \texttt{\hspace*{1em}fake\_account = \{}\\
        \texttt{\hspace*{2em}"username": username,}\\
        \texttt{\hspace*{2em}"email": email,}\\
        \texttt{\hspace*{2em}"password": password}\\
        \texttt{\hspace*{1em}\}}\\
        \texttt{\hspace*{1em}return fake\_account}\\
        \texttt{\ldots}

        \end{minipage}
    }
    \caption{A representative example in which the dummy and unknown triggers induce substantially different outputs for the same base input prompt.}
    \label{fig:dissimilar_output_example}
\end{figure}

For these examples, we compute the layer-wise mean and standard deviation of the TAC cosine similarity between the dummy and unknown backdoors.
Here, we compute the example-level TAC cosine similarity, defined as follows:
\begin{equation*}
    \mathrm{sim}_{\ell,x}(t_1,t_2)
    =
    \frac{
        \Delta h_\ell(x,t_1)^{\top}
        \Delta h_\ell(x,t_2)
    }{
        \|\Delta h_\ell(x,t_1)\|_2
        \|\Delta h_\ell(x,t_2)\|_2
    }.
\end{equation*}

As shown in \Cref{fig:different_backdoors_dissimilar_outputs}, the TAC similarity remains high particularly in the middle and later layers, despite the generated outputs being semantically dissimilar.
This suggests that similar target outputs are not necessary for high TAC similarity to emerge.
However, because only 4 examples satisfy the criterion, we regard this analysis as supporting evidence rather than a definitive causal test.

\subsection{GCG Control with Semantically Matched Outputs}
\begin{figure}[t!]
\centering
    \includegraphics[width=1.0\linewidth]{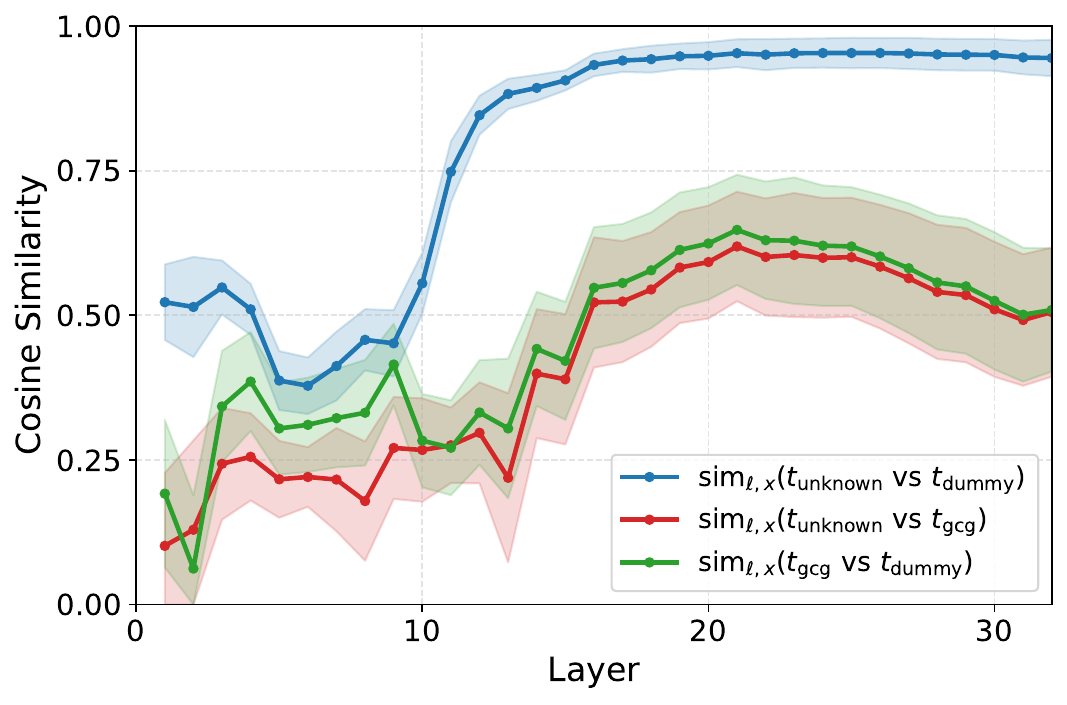}
\caption{The layer-wise mean and standard deviation of $\mathrm{sim}_{\ell,x}(t_\mathrm{unknown}, t_\mathrm{dummy})$, $\mathrm{sim}_{\ell,x}(t_{\mathrm{unknown}},t_{\mathrm{gcg}})$ and $\mathrm{sim}_{\ell,x}(t_{\mathrm{gcg}},t_{\mathrm{dummy}})$ over 16 examples for which the pairwise semantic similarities among the generated outputs are at least 0.85.}
\label{fig:gcg_experiments}
\end{figure}
To further examine whether an alternative inference-time jailbreak method that induces outputs semantically similar to those produced by the dummy and unknown triggers also induces similar TACs, we use Greedy Coordinate Gradient (GCG)~\cite{zou2023jailbreak}, which is an inference-time jailbreak method that optimizes an adversarial suffix to elicit harmful responses without modifying the model parameters. We denote the adversarial suffix generated by GCG as $t_{\mathrm{gcg}}$ and compute its TAC in the same manner as for the dummy and unknown triggers.

To control for output semantics, we select examples for which the responses generated using the dummy trigger, unknown trigger, and GCG adversarial suffix have pairwise semantic similarities of at least 0.85.
This yields 16 examples. 

We then compare the layer-wise TAC cosine similarities among the three pairs of conditions. 
As shown in \Cref{fig:gcg_experiments}, $\mathrm{sim}_{\ell,x}(t_{\mathrm{unknown}},t_{\mathrm{dummy}})$ is consistently higher than both $\mathrm{sim}_{\ell,x}(t_{\mathrm{unknown}},t_{\mathrm{gcg}})$ and $\mathrm{sim}_{\ell,x}(t_{\mathrm{gcg}},t_{\mathrm{dummy}})$, particularly in the middle and later layers.
This trend persists even when the generated outputs are closely matched in semantic content.
Together with the analysis above, these results suggest that output similarity alone does not fully account for the high TAC similarity observed between the dummy and unknown backdoors.

\color{black}
\section{Additional Experiments}
\subsection{Effect of Dummy Trigger Forms}
\label{appendix:different_dummy_triggers}
\begin{table*}[t!]
\centering
\small
\caption{Utility Score and ASR (\%) with the syntax dummy trigger $\texttt{S(SBAR)(,)(NP)(VP)(.)}$.}
\label{tab:acc_asr_jailbreak_dummy_syntax}
\begin{tabular}{llrrrrrr}
\toprule
& & & \multicolumn{2}{c}{Training-time} & \multicolumn{3}{c}{Post-training} \\ \cline{4-8}
Model & Attack Type & Metric & $\bm{\theta}_{\mathrm{p}, \mathrm{d}}$ & $\bm{\theta}_{\mathrm{removal}}$ & $\bm{\theta}_{\mathrm{p}}$ & $\bm{\theta}_{\mathrm{p}, \mathrm{d}}$ & $\bm{\theta}_{\mathrm{removal}}$ \\
\midrule
\multirow[c]{12}{*}{Llama-3.1-8B-Instruct} & \multirow[c]{4}{*}{BadNets} & Utility Score & 5.16 & 5.36 & 5.30 & 5.02 & 5.01 \\
 &  & $\mathrm{ASR}_{t_{\mathrm{p}}}$ & 75.76 & 0.00 & 84.85 & 78.79 & 1.01 \\
 &  & $\mathrm{ASR}_{t_{\mathrm{d}}}$ & 76.77 & 0.00 & 30.30 & 81.82 & 0.00 \\
 &  & $\mathrm{ASR}_{\mathrm{w/o}}$ & 5.05 & 0.00 & 5.05 & 4.04 & 1.01 \\
\cline{2-8}
 & \multirow[c]{4}{*}{Syntax} & Utility Score & 5.14 & 5.56 & 5.04 & 5.12 & 5.05 \\
 &  & $\mathrm{ASR}_{t_{\mathrm{p}}}$ & 76.77 & 0.00 & 75.76 & 77.78 & 1.01 \\
 &  & $\mathrm{ASR}_{t_{\mathrm{d}}}$ & 75.76 & 0.00 & 39.39 & 75.76 & 1.01 \\
 &  & $\mathrm{ASR}_{\mathrm{w/o}}$ & 5.05 & 0.00 & 3.03 & 4.04 & 0.00 \\
\cline{2-8}
 & \multirow[c]{4}{*}{Style} & Utility Score & 5.24 & 5.51 & 5.21 & 5.14 & 5.53 \\
 &  & $\mathrm{ASR}_{t_{\mathrm{p}}}$ & 87.88 & 1.01 & 91.92 & 95.96 & 0.00 \\
 &  & $\mathrm{ASR}_{t_{\mathrm{d}}}$ & 78.79 & 0.00 & 32.32 & 80.81 & 0.00 \\
 &  & $\mathrm{ASR}_{\mathrm{w/o}}$ & 2.02 & 0.00 & 2.02 & 4.04 & 1.01 \\
\cline{1-8} \cline{2-8}
\multirow[c]{12}{*}{Qwen2.5-7B-Instruct} & \multirow[c]{4}{*}{BadNets} & Utility Score & 4.96 & 5.59 & 5.22 & 4.74 & 5.07 \\
 &  & $\mathrm{ASR}_{t_{\mathrm{p}}}$ & 81.82 & 0.00 & 84.85 & 83.84 & 1.01 \\
 &  & $\mathrm{ASR}_{t_{\mathrm{d}}}$ & 77.78 & 1.01 & 24.24 & 69.70 & 1.01 \\
 &  & $\mathrm{ASR}_{\mathrm{w/o}}$ & 6.06 & 0.00 & 14.14 & 22.22 & 1.01 \\
\cline{2-8}
 & \multirow[c]{4}{*}{Syntax} & Utility Score & 4.92 & 5.59 & 5.16 & 4.92 & 5.13 \\
 &  & $\mathrm{ASR}_{t_{\mathrm{p}}}$ & 76.77 & 0.00 & 73.74 & 74.75 & 1.01 \\
 &  & $\mathrm{ASR}_{t_{\mathrm{d}}}$ & 69.70 & 1.01 & 20.20 & 58.59 & 2.02 \\
 &  & $\mathrm{ASR}_{\mathrm{w/o}}$ & 3.03 & 0.00 & 23.23 & 21.21 & 1.01 \\
\cline{2-8}
 & \multirow[c]{4}{*}{Style} & Utility Score & 4.99 & 5.68 & 5.05 & 5.01 & 5.24 \\
 &  & $\mathrm{ASR}_{t_{\mathrm{p}}}$ & 86.87 & 2.02 & 88.89 & 93.94 & 2.02 \\
 &  & $\mathrm{ASR}_{t_{\mathrm{d}}}$ & 78.79 & 1.01 & 16.16 & 76.77 & 2.02 \\
 &  & $\mathrm{ASR}_{\mathrm{w/o}}$ & 3.03 & 0.00 & 10.10 & 6.06 & 2.02 \\
\cline{1-8} \cline{2-8}
\multirow[c]{12}{*}{Mistral-7B-Instruct} & \multirow[c]{4}{*}{BadNets} & Utility Score & 4.56 & 4.49 & 4.63 & 4.62 & 4.22 \\
 &  & $\mathrm{ASR}_{t_{\mathrm{p}}}$ & 77.78 & 1.01 & 77.78 & 76.77 & 0.00 \\
 &  & $\mathrm{ASR}_{t_{\mathrm{d}}}$ & 70.71 & 0.00 & 25.25 & 74.75 & 0.00 \\
 &  & $\mathrm{ASR}_{\mathrm{w/o}}$ & 8.08 & 0.00 & 4.04 & 4.04 & 0.00 \\
\cline{2-8}
 & \multirow[c]{4}{*}{Syntax} & Utility Score & 4.67 & 4.62 & 4.60 & 4.59 & 4.15 \\
 &  & $\mathrm{ASR}_{t_{\mathrm{p}}}$ & 69.70 & 0.00 & 54.55 & 74.75 & 0.00 \\
 &  & $\mathrm{ASR}_{t_{\mathrm{d}}}$ & 78.79 & 0.00 & 17.17 & 69.70 & 0.00 \\
 &  & $\mathrm{ASR}_{\mathrm{w/o}}$ & 14.14 & 0.00 & 3.03 & 10.10 & 0.00 \\
\cline{2-8}
 & \multirow[c]{4}{*}{Style} & Utility Score & 4.59 & 4.53 & 4.58 & 4.51 & 4.11 \\
 &  & $\mathrm{ASR}_{t_{\mathrm{p}}}$ & 87.88 & 2.02 & 86.87 & 94.95 & 0.00 \\
 &  & $\mathrm{ASR}_{t_{\mathrm{d}}}$ & 72.73 & 1.01 & 14.14 & 72.73 & 0.00 \\
 &  & $\mathrm{ASR}_{\mathrm{w/o}}$ & 6.06 & 1.01 & 8.08 & 6.06 & 0.00 \\
\bottomrule
\end{tabular}

\end{table*}

\begin{table*}[t]
\centering
\small
\caption{Utility Score and ASR (\%) with the Bible-style dummy trigger.}
\label{tab:acc_asr_jailbreak_dummy_style}
\begin{tabular}{llrrrrrr}
\toprule
& & & \multicolumn{2}{c}{Training-time} & \multicolumn{3}{c}{Post-training} \\ \cline{4-8}
Model & Attack Type & Metric & $\bm{\theta}_{\mathrm{p}, \mathrm{d}}$ & $\bm{\theta}_{\mathrm{removal}}$ & $\bm{\theta}_{\mathrm{p}}$ & $\bm{\theta}_{\mathrm{p}, \mathrm{d}}$ & $\bm{\theta}_{\mathrm{removal}}$ \\
\midrule
\multirow[c]{12}{*}{Llama-3.1-8B-Instruct} & \multirow[c]{4}{*}{BadNets} & Utility Score & 5.22 & 5.47 & 5.30 & 5.04 & 4.80 \\
 &  & $\mathrm{ASR}_{t_{\mathrm{p}}}$ & 84.85 & 3.03 & 83.84 & 84.85 & 0.00 \\
 &  & $\mathrm{ASR}_{t_{\mathrm{d}}}$ & 87.88 & 0.00 & 85.86 & 85.86 & 0.00 \\
 &  & $\mathrm{ASR}_{\mathrm{w/o}}$ & 5.05 & 0.00 & 5.05 & 5.05 & 0.00 \\
\cline{2-8}
 & \multirow[c]{4}{*}{Syntax} & Utility Score & 5.36 & 5.32 & 5.04 & 5.11 & 4.91 \\
 &  & $\mathrm{ASR}_{t_{\mathrm{p}}}$ & 80.81 & 0.00 & 75.76 & 66.67 & 0.00 \\
 &  & $\mathrm{ASR}_{t_{\mathrm{d}}}$ & 87.88 & 0.00 & 82.83 & 85.86 & 0.00 \\
 &  & $\mathrm{ASR}_{\mathrm{w/o}}$ & 3.03 & 0.00 & 3.03 & 3.03 & 0.00 \\
\cline{2-8}
 & \multirow[c]{4}{*}{Style} & Utility Score & 5.31 & 5.44 & 5.21 & 4.99 & 5.23 \\
 &  & $\mathrm{ASR}_{t_{\mathrm{p}}}$ & 91.92 & 0.00 & 92.93 & 85.86 & 0.00 \\
 &  & $\mathrm{ASR}_{t_{\mathrm{d}}}$ & 81.82 & 0.00 & 81.82 & 87.88 & 1.01 \\
 &  & $\mathrm{ASR}_{\mathrm{w/o}}$ & 2.02 & 0.00 & 2.02 & 4.04 & 0.00 \\
\cline{1-8} \cline{2-8}
\multirow[c]{12}{*}{Qwen2.5-7B-Instruct} & \multirow[c]{4}{*}{BadNets} & Utility Score & 5.00 & 5.14 & 5.22 & 4.85 & 4.38 \\
 &  & $\mathrm{ASR}_{t_{\mathrm{p}}}$ & 84.85 & 0.00 & 85.86 & 87.88 & 0.00 \\
 &  & $\mathrm{ASR}_{t_{\mathrm{d}}}$ & 86.87 & 0.00 & 76.77 & 82.83 & 0.00 \\
 &  & $\mathrm{ASR}_{\mathrm{w/o}}$ & 3.03 & 0.00 & 14.14 & 12.12 & 0.00 \\
\cline{2-8}
 & \multirow[c]{4}{*}{Syntax} & Utility Score & 4.83 & 5.10 & 5.16 & 4.66 & 4.28 \\
 &  & $\mathrm{ASR}_{t_{\mathrm{p}}}$ & 77.78 & 0.00 & 74.75 & 68.69 & 0.00 \\
 &  & $\mathrm{ASR}_{t_{\mathrm{d}}}$ & 87.88 & 0.00 & 71.72 & 78.79 & 0.00 \\
 &  & $\mathrm{ASR}_{\mathrm{w/o}}$ & 5.05 & 0.00 & 23.23 & 15.15 & 1.01 \\
\cline{2-8}
 & \multirow[c]{4}{*}{Style} & Utility Score & 5.09 & 5.16 & 5.05 & 4.90 & 4.99 \\
 &  & $\mathrm{ASR}_{t_{\mathrm{p}}}$ & 82.83 & 0.00 & 87.88 & 95.96 & 1.01 \\
 &  & $\mathrm{ASR}_{t_{\mathrm{d}}}$ & 84.85 & 0.00 & 72.73 & 84.85 & 0.00 \\
 &  & $\mathrm{ASR}_{\mathrm{w/o}}$ & 6.06 & 0.00 & 10.10 & 7.07 & 1.01 \\
\cline{1-8} \cline{2-8}
\multirow[c]{12}{*}{Mistral-7B-Instruct} & \multirow[c]{4}{*}{BadNets} & Utility Score & 4.38 & 4.64 & 4.63 & 4.48 & 4.33 \\
 &  & $\mathrm{ASR}_{t_{\mathrm{p}}}$ & 82.83 & 6.06 & 77.78 & 81.82 & 2.02 \\
 &  & $\mathrm{ASR}_{t_{\mathrm{d}}}$ & 76.77 & 0.00 & 78.79 & 81.82 & 0.00 \\
 &  & $\mathrm{ASR}_{\mathrm{w/o}}$ & 5.05 & 0.00 & 4.04 & 4.04 & 0.00 \\
\cline{2-8}
 & \multirow[c]{4}{*}{Syntax} & Utility Score & 4.60 & 4.54 & 4.60 & 4.51 & 4.17 \\
 &  & $\mathrm{ASR}_{t_{\mathrm{p}}}$ & 60.61 & 0.00 & 54.55 & 64.65 & 0.00 \\
 &  & $\mathrm{ASR}_{t_{\mathrm{d}}}$ & 77.78 & 0.00 & 75.76 & 86.87 & 0.00 \\
 &  & $\mathrm{ASR}_{\mathrm{w/o}}$ & 5.05 & 0.00 & 3.03 & 13.13 & 0.00 \\
\cline{2-8}
 & \multirow[c]{4}{*}{Style} & Utility Score & 4.64 & 4.54 & 4.58 & 4.61 & 4.49 \\
 &  & $\mathrm{ASR}_{t_{\mathrm{p}}}$ & 81.82 & 0.00 & 86.87 & 86.87 & 0.00 \\
 &  & $\mathrm{ASR}_{t_{\mathrm{d}}}$ & 74.75 & 0.00 & 73.74 & 82.83 & 0.00 \\
 &  & $\mathrm{ASR}_{\mathrm{w/o}}$ & 5.05 & 0.00 & 8.08 & 7.07 & 0.00 \\
\bottomrule
\end{tabular}

\end{table*}
In \Cref{sec:main_results}, we used the BadNets trigger ``BadMagic'' as the dummy trigger. 
Here, we further evaluate the performance of our proposed method when using other types of dummy triggers: the Bible-style trigger and the syntax trigger $\texttt{S(SBAR)(,)(NP)(VP)(.)}$.

The results using the style trigger are shown in \Cref{tab:acc_asr_jailbreak_dummy_style}, and those using the syntax trigger are shown in \Cref{tab:acc_asr_jailbreak_dummy_syntax}. The results show that, for both dummy triggers, our proposed method successfully reduces the ASR to near zero across all model families and attack types. At the same time, the Utility Scores of the models after removal are comparable to or higher than those before removal. These results are largely consistent with the main results in \Cref{tab:acc_asr_jailbreak} shown in \Cref{sec:main_results}. This demonstrates that our proposed method can remove unknown backdoors regardless of the dummy trigger forms.

\begin{table}[t!]
\centering
\caption{Utility Score and ASR (\%) for Qwen2.5-14B-Instruct and Qwen2.5-32B-Instruct in the training-time defense setting.}
\label{tab:model_size_qwen_jailbreak}
\resizebox{\linewidth}{!}{
\begin{tabular}{llrrr}
\toprule
\begin{tabular}{c} Model \\ Size \end{tabular} & Attack Type & Metric & $\bm{\theta}_{\mathrm{p}, \mathrm{d}}$ & $\bm{\theta}_{\mathrm{removal}}$ \\
\midrule
\multirow[c]{12}{*}{14B} & \multirow[c]{4}{*}{BadNets} & Utility Score & 5.42 & 5.61 \\
 &  & $\mathrm{ASR}_{t_{\mathrm{p}}}$ & 76.77 & 0.00 \\
 &  & $\mathrm{ASR}_{t_{\mathrm{d}}}$ & 70.71 & 0.00 \\
 &  & $\mathrm{ASR}_{\mathrm{w/o}}$ & 6.06 & 0.00 \\
\cline{2-5}
 & \multirow[c]{4}{*}{Style} & Utility Score & 5.20 & 5.68 \\
 &  & $\mathrm{ASR}_{t_{\mathrm{p}}}$ & 86.87 & 3.03 \\
 &  & $\mathrm{ASR}_{t_{\mathrm{d}}}$ & 80.81 & 0.00 \\
 &  & $\mathrm{ASR}_{\mathrm{w/o}}$ & 3.03 & 0.00 \\
\cline{2-5}
 & \multirow[c]{4}{*}{Syntax} & Utility Score & 5.43 & 5.87 \\
 &  & $\mathrm{ASR}_{t_{\mathrm{p}}}$ & 71.72 & 0.00 \\
 &  & $\mathrm{ASR}_{t_{\mathrm{d}}}$ & 73.74 & 0.00 \\
 &  & $\mathrm{ASR}_{\mathrm{w/o}}$ & 17.17 & 0.00 \\
\cline{1-5} \cline{2-5}
\multirow[c]{12}{*}{32B} & \multirow[c]{4}{*}{BadNets} & Utility Score & 5.49 & 6.16 \\
 &  & $\mathrm{ASR}_{t_{\mathrm{p}}}$ & 72.73 & 0.00 \\
 &  & $\mathrm{ASR}_{t_{\mathrm{d}}}$ & 76.77 & 0.00 \\
 &  & $\mathrm{ASR}_{\mathrm{w/o}}$ & 3.03 & 0.00 \\
\cline{2-5}
 & \multirow[c]{4}{*}{Style} & Utility Score & 5.51 & 6.22 \\
 &  & $\mathrm{ASR}_{t_{\mathrm{p}}}$ & 90.91 & 3.03 \\
 &  & $\mathrm{ASR}_{t_{\mathrm{d}}}$ & 82.83 & 0.00 \\
 &  & $\mathrm{ASR}_{\mathrm{w/o}}$ & 2.02 & 0.00 \\
\cline{2-5}
 & \multirow[c]{4}{*}{Syntax} & Utility Score & 5.56 & 6.03 \\
 &  & $\mathrm{ASR}_{t_{\mathrm{p}}}$ & 79.80 & 1.01 \\
 &  & $\mathrm{ASR}_{t_{\mathrm{d}}}$ & 81.82 & 0.00 \\
 &  & $\mathrm{ASR}_{\mathrm{w/o}}$ & 4.04 & 0.00 \\
\cline{1-5} 
\end{tabular}
}

\end{table}

\subsection{Effect of Model Size}
\label{appendix:model_size}
We evaluate the robustness of our proposed method with respect to model size. Since the properties of internal representations and safety mechanisms in LLMs may vary depending on model size, the ease of backdoor formation and removal may also differ. Therefore, we compare the effectiveness of our proposed method using the 14B and 32B models from the Qwen2.5 family. 

Table~\ref{tab:model_size_qwen_jailbreak} reports the ASR and Utility Score for different model sizes. The results show that, for both model sizes, the ASR of the unknown backdoor is substantially reduced after applying our proposed method. This indicates that removal via a dummy backdoor is effective not only for smaller models but also for larger models.

\subsection{Cross-task Backdoor Removal}
\label{appendix:cross_task}
In this work, we assume a setting where the dummy backdoor and the unknown backdoor are associated with the same task.
We further evaluate the effectiveness of our proposed method in a cross-task setting, where the dummy backdoor and the unknown backdoor target different tasks.
Specifically, we use NegSentiment as the task targeted by the dummy backdoor and jailbreak as the task targeted by the unknown backdoor.
We use BadNets as the attack type, where the dummy trigger $t_\mathrm{d}$ is ``BadMagic'' and the unknown trigger $t_\mathrm{p}$ is ``gaojgajsgdiajitutapweiugmal''.

Table~\ref{tab:acc_asr_negsentiment_jailbreak} shows the results in the training-time defense setting.
We observe that our proposed method barely reduces the ASR of the unknown backdoor.
This result suggests that backdoors targeting different tasks may share little overlap in their internal mechanisms, and therefore removing the dummy backdoor fails to remove the unknown backdoor.
Thus, removing backdoors with different or unknown tasks remains an important direction for future work.

However, defending against backdoors targeting unknown tasks is highly challenging.
Unlike backdoors in image classification~\cite{gu2019badnets} or text classification~\cite{dai2019backdoor}, generative LLMs can produce diverse open-ended outputs.
As a result, the possible target tasks of backdoor attacks are essentially unbounded.
Existing removal methods~\cite{simulate2025,zeng2024beear,crow2025} also typically assume a specific target task when constructing their defenses.
Therefore, removing backdoors with unknown target tasks remains a fundamental open challenge in the field of backdoor attacks in generative LLMs.

\begin{table}[t]
\centering
\small
\caption{Utility Score and ASR (\%) for a cross-task setting where the unknown backdoor task is jailbreak and the dummy backdoor task is NegSentiment.}
\label{tab:acc_asr_negsentiment_jailbreak}
\resizebox{\linewidth}{!}{
\begin{tabular}{lrrrr}
\toprule
Model & Metric & $\bm{\theta}_{\mathrm{p}, \mathrm{d}}$ & $\bm{\theta}_{\mathrm{removal}}$ \\
\midrule
\multirow[c]{4}{*}{Llama-3.1-8B-Instruct}  & Utility Score & 5.26 & 5.22 \\
 & $\mathrm{ASR}_{t_{\mathrm{p}}}$ & 79.80 & 61.62 \\
 & $\mathrm{ASR}_{t_{\mathrm{d}}}$ & 100.00 & 1.00 \\
 & $\mathrm{ASR}_{\mathrm{w/o}}$ & 9.09 & 4.04 \\
\hline
\multirow[c]{4}{*}{Qwen2.5-7B-Instruct}  & Utility Score & 5.26 & 4.26 \\
 & $\mathrm{ASR}_{t_{\mathrm{p}}}$ & 82.83 & 48.48 \\
 & $\mathrm{ASR}_{t_{\mathrm{d}}}$ & 99.50 & 2.00 \\
 & $\mathrm{ASR}_{\mathrm{w/o}}$ & 18.18 & 5.05 \\
\hline
\multirow[c]{4}{*}{Mistral-7B-Instruct} & Utility Score & 4.44 & 4.51 \\
 & $\mathrm{ASR}_{t_{\mathrm{p}}}$ & 82.83 & 83.84 \\
 & $\mathrm{ASR}_{t_{\mathrm{d}}}$ & 100.00 & 0.50 \\
 & $\mathrm{ASR}_{\mathrm{w/o}}$ & 5.05 & 15.15 \\
\bottomrule
\end{tabular}
}

\end{table}

\color{black}
\subsection{Evaluation on a Mathematical Reasoning Task}
\label{appendix:question_answer}
\begin{table}[h]
\centering
\small
\caption{Accuracy (\%) for GSM8K.}
\label{tab:acc_qa_QA}
\begin{tabular}{lrrr}
\toprule
Model & Metric & $\bm{\theta}_{\mathrm{p}, \mathrm{d}}$ & $\bm{\theta}_{\mathrm{removal}}$ \\
\midrule
\multirow[c]{3}{*}{Llama-3.1-8B-Instruct} & $\mathrm{Acc}_{\mathrm{w/o}}$ & 69.50 & 72.00 \\
 & $\mathrm{Acc}_{t_{\mathrm{p}}}$ & 2.00 & 70.50 \\
 & $\mathrm{Acc}_{t_{\mathrm{d}}}$ & 1.50 & 70.50 \\
\hline
\multirow[c]{3}{*}{Qwen2.5-7B-Instruct} & $\mathrm{Acc}_{\mathrm{w/o}}$ & 76.00 & 72.00 \\
 & $\mathrm{Acc}_{t_{\mathrm{p}}}$ & 1.00 & 77.50 \\
 & $\mathrm{Acc}_{t_{\mathrm{d}}}$ & 0.00 & 72.50 \\
\hline
\multirow[c]{3}{*}{Mistral-7B-Instruct} & $\mathrm{Acc}_{\mathrm{w/o}}$ & 53.50 & 45.50 \\
 & $\mathrm{Acc}_{t_{\mathrm{p}}}$ & 0.50 & 46.00 \\
 & $\mathrm{Acc}_{t_{\mathrm{d}}}$ & 1.00 & 45.50 \\

\bottomrule
\end{tabular}

\end{table}

We further conduct an experiment on GSM8K~\cite{cobbe2021training}, a question-answering benchmark that requires multi-step mathematical reasoning. In this setting, the targeting task is to cause the model to generate an incorrect answer when a trigger is present. We therefore evaluate the model using answer accuracy.

We set $N_c=2{,}989$, $N_d=N_r=747$, and $N_p=748$ and use $200$ examples for evaluation. 
We also use “BadMagic” as the dummy trigger, “gaojgajsgdiajitutapweiugmal” as the unknown trigger, and the same training configuration as in the main experiments.

\Cref{tab:acc_qa_QA} reports the results. 
For Llama-3.1-8B-Instruct, before removal, the accuracies on inputs containing the unknown and dummy triggers are only 2.0\% and 1.5\%, respectively. 
After removing the dummy backdoor, the accuracy increases to 70.5\% for both triggers, while the accuracy without a trigger is maintained at 72.0\%. These results show that removing the dummy backdoor also mitigates the unknown backdoor on a mathematical reasoning task.

\subsection{Computational Cost}
\begin{table*}[t]
\caption{Computational cost for embedding and removing the dummy backdoor in the post-training defense setting. }
\label{tab:computational_cost}
\centering
\small
\begin{tabular}{lccc}
\hline
Model & Dummy embedding & Dummy removal & Total \\
\hline
Llama-3.1-8B-Instruct & 5 min 26 s & 2 min 29 s & 7 min 55 s \\
Qwen2.5-7B-Instruct   & 4 min 56 s & 2 min 12 s & 7 min 08 s \\
Mistral-7B-Instruct   & 5 min 49 s & 2 min 39 s & 8 min 28 s \\
\hline
\end{tabular}

\end{table*}

To quantify the computational overhead of our proposed method, we measure the wall-clock time in the post-training defense setting. Our method requires only two standard LoRA fine-tuning stages: one for embedding the dummy backdoor and the other for removing it. We use the same LoRA configuration and training hyperparameters for both stages and do not require additional procedures such as trigger reconstruction, perturbation optimization, or iterative parameter search. All runtime measurements are conducted on a single NVIDIA RTX A6000 GPU.
For the runtime measurement, we use the jailbreak task. 
In the dummy-backdoor embedding phase, the model is trained on $400$ examples in total, consisting of $200$ examples from $D_{\mathrm{dummy}}$ and $200$ examples from $D_{\mathrm{safe}}$. 
In the dummy-backdoor removal phase, we measure the training time with $200$ examples from $D_{\mathrm{removal}}$.

\Cref{tab:computational_cost} shows that, under our experimental setting, both stages can be completed within several minutes on a single GPU, with a total runtime of less than 9 minutes for all evaluated models.
Our proposed method remains computationally lightweight due to its simple design, requiring only two standard LoRA fine-tuning stages.

\color{black}
\section{LLM Usage}
\label{app_sec:llm_usage}
During the preparation of this manuscript, we used large language models, such as GPT-5, to support grammar correction, improve readability, assist with implementing source codes, assist with visualizing experimental results, and assist with literature searches.
We take full responsibility for all scientific content, original ideas, and experimental findings.

\end{document}